\documentclass[a4paper, 11pt, oneside]{article}
\pdfoutput=1
\usepackage{float}
\usepackage{jheppub}
\usepackage{epstopdf}
\usepackage{mathtools}
\usepackage{enumitem}

\usepackage{bbm}   
\usepackage{amsmath}
\usepackage{graphicx}
\usepackage{ulem}
\usepackage{mathrsfs}
\usepackage{xcolor}

\newcommand{\nc}{\newcommand}

\newlength{\absize}
\nc{\non}{\nonumber}
\nc{\hc}{\hbox {H.c.}} 
\nc{\noi}{\noindent}
\nc{\barx}{\bar{x}}
\nc{\pbarn}{\;\hbox {pb}}
\nc{\fbarn}{\;\hbox {fb}}
\newcommand{\bi}{\begin{itemize}}
\newcommand{\ei}{\end{itemize}}

\def\thetaW{{\theta}_\text{W}}
\nc{\ssp}{\;\;\;}
\nc{\lsp}{\;\;\;\;\;}
\nc{\Lsp}{\;\;\;\;\;\;\;\;\;\;}  
\nc{\LLsp}{\lspace \lspace}
\nc{\lra}{\longrightarrow}
\def\ifmath#1{\relax\ifmmode #1\else $#1$\fi}
\newenvironment{Eqnarray}%
         {\arraycolsep 0.14em\begin{eqnarray}}{\end{eqnarray}}
\nc{\beq}{\begin{equation}}  \nc{\eeq}{\end{equation}}
\nc{\bea}{\begin{Eqnarray}}  \nc{\eea}{\end{Eqnarray}}
\nc{\baa}{\begin{array}}     \nc{\eaa}{\end{array}}
\nc{\bit}{\begin{itemize}}   \nc{\eit}{\end{itemize}}
\nc{\ben}{\begin{enumerate}} \nc{\een}{\end{enumerate}}
\nc{\bce}{\begin{center}}    \nc{\ece}{\end{center}}
\nc{\bpm}{\begin{pmatrix}}   \nc{\epm}{\end{pmatrix}}
\nc{\bvt}{\begin{verbatim}}  \nc{\evt}{\end{verbatim}}
\def\lsim{\mathrel{\raise.3ex\hbox{$<$\kern-.75em\lower1ex\hbox{$\sim$}}}}
\def\gsim{\mathrel{\raise.3ex\hbox{$>$\kern-.75em\lower1ex\hbox{$\sim$}}}}
\def\eq#1{eq.~(\ref{#1})}

\def\eqs#1#2{eqs.~(\ref{#1}) and (\ref{#2})}
\def\Eq#1{Eq.~(\ref{#1})}
\def\Eqs#1#2{Eqs.~(\ref{#1}) and (\ref{#2})}

\def\nn{\nonumber}
\renewcommand{\Re}{\mbox{Re\thinspace}}
\renewcommand{\Im}{\mbox{Im\thinspace}}

\newcommand{\half}{{\textstyle\frac{1}{2}}}
\def\thetaW{{\theta}_\text{W}}
\newcommand{\BR}{\text{BR}}
 \def\ddel{\!\!\mathrel{\raise1.5ex\hbox{$\leftrightarrow$\kern-.85em
\lower1.7ex\hbox{$\partial$}}}}

\def\hcal{{\cal H}}

\def\lcal{{\cal L}}

\def\pcal{{\cal P}}

\def\gev{\;\hbox{GeV}}
\def\tev{\;\hbox{TeV}}
\def\zBB{{\mathbbm Z}}
\def\z2{\zBB_2}
\def\phaa{\phantom{AA}}

\nc{\tanb}{\tan\beta}
\nc{\mch}{M_{H^\pm}}

\def\cb{c_\beta}
\def\sb{s_\beta}

\def\mch{M_{H^\pm}}
\def\lsub#1{\ifmath{_{\lower1.5pt\hbox{$\scriptstyle #1$}}}}
\def\lsup#1{^{\lower 6pt\hbox{$\scriptstyle#1$}}}

\def\ddel{\!\!\mathrel{\raise1.5ex\hbox{$\leftrightarrow$\kern-.85em
\lower1.7ex\hbox{$\partial$}}}}

\nc{\for}{\lsp {\rm for} \lsp}
\nc{\andd}{\lsp {\rm and} \lsp}

\renewcommand{\Re}{\mbox{Re\thinspace}}
\renewcommand{\Im}{\mbox{Im\thinspace}}

\allowdisplaybreaks %!!!!

\def\i11{{\mathbbm 1}}%\def\iBB{ \hbox{{\mysmallii I}}\!\hbox{{\mysmallii I}} }

\title{Heavy Higgs boson decays in the alignment limit of the 2HDM}
\author[a]{Bohdan Grzadkowski,}
\affiliation[a]{Faculty of Physics, University of Warsaw, Pasteura 5, 02-093 Warsaw, Poland}
\author[b]{Howard E. Haber,}
\affiliation[b]{Santa Cruz Institute for Particle Physics,
University of California,
1156 High Street,
Santa Cruz, California 95064, USA}
\author[c]{Odd Magne Ogreid,}
\affiliation[c]{Western Norway University of Applied Sciences,
Postboks 7030, N-5020 Bergen, Norway, }
\author[d]{Per Osland}
\affiliation[d]{Department of Physics,
University of Bergen, Postboks 7803, N-5020 Bergen, Norway}
\emailAdd{bohdan.grzadkowski@fuw.edu.pl}
\emailAdd{haber@scipp.ucsc.edu}
\emailAdd{omo@hvl.no}
\emailAdd{Per.Osland@uib.no}
\date{\today}

\abstract{The Standard Model (SM)-like couplings of the observed Higgs boson impose strong constraints on the structure of any extended Higgs sector.  
We consider the theoretical properties and the phenomenological implications of a generic two Higgs doublet model (2HDM).
This model constitutes a simple and attractive extension of the SM that is consistent with the observation of the SM-like Higgs boson and precision electroweak observables, while providing a potential new source of CP-violation. In this paper we focus on the so-called Higgs alignment limit of the generic 2HDM, where
the neutral scalar field~$H_1$,
with the tree-level couplings of the SM Higgs boson, is a mass eigenstate that is aligned in field space with the direction of the Higgs vacuum expectation value.  The properties of the two other heavier neutral Higgs scalars, $H_2$ and $H_3$, in the alignment limit of the 2HDM are also elucidated.
It is shown that the couplings of $H_2$ and $H_3$ in the alignment limit are tightly constrained and correlated. 
For example, in the exact alignment limit at tree level, for bosonic final states $\BR(H_{2,3} \to W^+W^-, ZZ, H_1 Z) = 0$ and
$\BR(H^\pm \to W^\pm H_1) = 0$, whereas for fermionic final states  $\Gamma(H_2 \to f\bar f)/\Gamma(H_3 \to f\bar f) \sim M_2/M_3$ (where $M_\alpha$ is the mass of $H_\alpha$).  In some cases, the results of the alignment limit differ depending on whether or not alignment is achieved via the decoupling of heavy scalar states.   In particular, in the exact alignment limit without decoupling BR($H_{2,3}\to H_1 H_1)=0$, whereas these branching ratios are nonzero in the decoupling regime.
Observables that could be used to test the alignment scenario at the LHC are defined and discussed.
The couplings of the Higgs bosons away from their exact alignment values are determined to leading order,
and some consequences are elucidated.}

\begin{document}
\vspace*{-1.5cm}
\begin{flushright}
SCIPP-18/05\\
\end{flushright}

\maketitle

\flushbottom

%%%%%%%%%%%%%%%%%%%%%%%%%%%%%%%%%%%%%%%%%%%%%%%%%%%%%%%%%%%%%
\section{Introduction}
\label{Sec:Introduction}
%%%%%%%%%%%%%%%%%%%%%%%%%%%%%%%%%%%%%%%%%%%%%%%%%%%%%%%%%%%%%
It is widely believed that the Standard Model (SM) of electroweak interactions is merely an effective theory valid up to an energy scale  of $\sim 200-300\gev$. New, heavier degrees of freedom may exist, and their discovery would be direct evidence for 
beyond the SM physics. Here we will focus on searches for new states that have spin zero, i.e., we are going to consider extensions of the scalar sector of the SM. If certain constraints known as the alignment limit (AL) are satisfied, then it turns out that the new scalars would not necessarily be much heavier than the discovered $125\gev$ Higgs boson. 
In order to discover modifications of the scalar sector both the ATLAS and CMS collaborations at the LHC are looking for new spin-zero resonances. These searches are aimed at different final-state channels, $t \bar t$ \cite{Sirunyan:2017uyt,Aaboud:2017hnm}, $b\bar b$ or lepton pairs ($\tau^+\tau^-$, $\mu^+\mu^-$) \cite{Sirunyan:2017isc,Sirunyan:2017uvf,Aaboud:2017sjh}, electroweak gauge bosons \cite{Aaboud:2017itg,Aaboud:2017gsl,Aaboud:2017fgj,Aaboud:2017rel}, diphoton states \cite{Aaboud:2017yyg} or an electroweak gauge boson in association with the SM Higgs boson \cite{Aaboud:2017cxo}.
In this context it is worth re-examining new physics beyond the SM that can arise due to an extended Higgs sector.

In light of the measured value of the electroweak $\rho$-parameter~\cite{Ross:1975fq} that is close to~$1$~\cite{Tanabashi:2018oca}, the most natural choice of an extended Higgs sector consists of scalar fields in singlet and doublet representations of the SU(2) gauge symmetry.  In this paper, we focus our attention on the two Higgs doublet extension of the SM (2HDM), as it is the most modest extension of the SM that contains a number of interesting new phenomena beyond the SM such as charged scalars and neutral scalars of potentially indefinite CP.   
The latter is a consequence of a new source of CP-violation (CPV) originating in the 2HDM scalar potential, which is required by a desire to explain the baryon asymmetry~\cite{White:2016nbo} observed in the Universe.

In the literature, much attention has been given to 2HDM Lagrangians that possess a $\mathbb{Z}_2$ symmetry (perhaps softly broken), which provides a natural mechanism for avoiding tree-level flavor-changing neutral currents mediated by neutral Higgs exchange~\cite{Glashow:1976nt,Paschos:1976ay}.  In this work, we relax this assumption to consider the most general 2HDM.   Of course, one must be careful to make sure that the parameter space of this model is consistent with all known experimental constraints.  These considerations imply the existence of two approximate alignments of 2HDM parameters.  First, the Higgs--fermion Yukawa couplings must be approximately flavor-aligned to ensure that flavor-changing neutral currents are sufficiently suppressed~\cite{Manohar:2006ga,Pich:2009sp,Bijnens:2011gd,Gori:2017qwg,Penuelas:2017ikk}.  Second, given the consistency of the Higgs precision data with SM predictions with an accuracy of approximately 20\%~\cite{ATLASHiggs,CMSHiggs}, the 2HDM parameters must be close to the values obtained in the Higgs alignment limit (AL).  In this limit, a scalar field that is aligned in field space with the Higgs vacuum expectation value (and therefore possesses the tree-level couplings of the SM Higgs field) is a mass eigenstate, which is identified with the observed Higgs boson of mass 125 GeV~\cite{Craig:2013hca,Haber:2013mia,Asner:2013psa,Carena:2013ooa} (see also Refs.~\cite{Dev:2014yca,Das:2015mwa,Bernon:2015qea,Bernon:2015wef}).  This latter constraint, when applied to the softly-broken $\mathbb{Z}_2$-symmetric version of the 2HDM, suppresses the possibility of new CPV phenomena originating in the scalar potential (e.g., as shown in Ref.~\cite{Grzadkowski:2014ada}).  Hence, we shall dispense of the $\mathbb{Z}_2$ symmetry and consider the most general 2HDM, subject only to the phenomenological constraints on its parameters. 
Moreover, in the complete absence of a $\mathbb{Z}_2$ symmetry (in the AL), new sources of CP-violation can arise both in the scalar potential and in  the Yukawa interactions of the neutral heavier Higgs bosons (which in this case are phenomenologically less constrained).

In the exact AL of the 2HDM, one Higgs boson (e.g., the lightest one, which is assumed in this work)
couples to vector bosons and fermions with tree-level couplings that are precisely those of the SM Higgs boson. 
However, in the case of alignment without decoupling, the heavier neutral ($H_{2,3}$) and charged
($H^\pm$) states can still be relatively light (with masses of order the electroweak scale), so that they can be detected and studied at the LHC.
The goal of this paper is to investigate interactions of the heavy scalars in the AL.
It turns out that in the AL properties of $H_2$ and $H_3$  are strongly correlated, which implies various relations between
observables involving $H_2$ and $H_3$. First, we are going to determine the correlations between $H_{2,3}$ couplings. Next, we define observables which could test the alignment scenario. Then, whenever possible, we will try to suggest measurements that
can disentangle the different types of Yukawa couplings in the 2HDM,
Type I, Type II and generic Yukawa couplings, assuming the AL. 

Of course, the LHC data are subject to uncertainties and therefore a dedicated numerical analysis in the vicinity of the AL is mandatory. Nevertheless, we 
believe that the study of the heavy Higgs bosons in the exact AL provides a natural guidance and should be helpful for  experimental searches for heavy Higgs bosons.

This work is organized as follows. 
After presenting the motivation for this work,
in section~\ref{Sec:def-model} we introduce the model and necessary notation.
In section~\ref{Sec:input} we specify the input parameters and discuss the issue of decoupling versus alignment in the 2HDM. 
Section~\ref{Sec:alignment} is devoted to the alignment limit of the model.
The extra freedom provided by the generic 2HDM in the AL is illustrated by gluon fusion in section~\ref{Sec:production}.
Decays of extra Higgs bosons in the generic 2HDM in the AL are discussed in section~\ref{Heavy_Higgs_decays} with an emphasis on correlations between various decays.
Appendices contain a comprehensive list of Higgs boson couplings in the generic 2HDM and some of its most popular versions.

%%%%%%%%%%%%%%%%%%%%%%%%%%%%%%%%%%%%%%%%%%%%%%%%%%%%%%%%%%%%
\section{The model}
\label{Sec:def-model}
\setcounter{equation}{0}

The scalar potential of the 2HDM shall be parametrized in the standard fashion:
\bea
\!\!\!\!\!\!\!\!\!\!\!\!\!\!\!
V(\Phi_1,\Phi_2) &=& -\tfrac12\left\{m_{11}^2\Phi_1^\dagger\Phi_1
+ m_{22}^2\Phi_2^\dagger\Phi_2 + \left[m_{12}^2 \Phi_1^\dagger \Phi_2
+ \hc\right]\right\} \nonumber \\
&& + \tfrac12\lambda_1(\Phi_1^\dagger\Phi_1)^2
+ \tfrac12\lambda_2(\Phi_2^\dagger\Phi_2)^2
+ \lambda_3(\Phi_1^\dagger\Phi_1)(\Phi_2^\dagger\Phi_2) 
+ \lambda_4(\Phi_1^\dagger\Phi_2)(\Phi_2^\dagger\Phi_1) \nn
\\[1pt]
&&+ \left\{\tfrac12\lambda_5(\Phi_1^\dagger\Phi_2)^2 
+\bigl[\lambda_6(\Phi_1^\dagger\Phi_1)+\lambda_7
(\Phi_2^\dagger\Phi_2)\bigr](\Phi_1^\dagger\Phi_2)
+{\rm \hc}\right\} .\label{Eq:pot}
\eea

Usually a $\zBB_2$ symmetry is imposed on the dimension-4 terms in order to eliminate potentially large flavor-changing neutral currents in the Yukawa couplings. In the present work we will not restrict ourselves by imposing this symmetry. Instead, we are going to consider the most
general scalar potential, keeping also terms that are not allowed by $\zBB_2$ symmetry. We will refer to this model as the 2HDM67, emphasizing the presence of $\lambda_6$ and $\lambda_7$ in the potential. 

In a general basis, the vacuum may be complex, and the Higgs doublets shall be parametrized as 
\begin{equation}
\Phi_j=e^{i\xi_j}\left(
\begin{array}{c}\varphi_j^+\\ (v_j+\eta_j+i\chi_j)/\sqrt{2}
\end{array}\right), \quad
j=1,2,\label{vevs}
\end{equation}
with the $v_j$ real numbers, satisfying $v_1^2+v_2^2=v^2$. The fields $\eta_j$ and $\chi_j$ are real,
and the difference between the phases of the two vacuum expectation values (VEVs) is denoted by
\beq
\xi\equiv\xi_2-\xi_1.
\eeq
Next, we shall define orthogonal states
\beq
\left(
\begin{array}{c}G^0\\ \eta_3
\end{array}\right)
=
\left(
\begin{array}{cc} v_1/v & \quad v_2/v\\ -v_2/v &\quad  v_1/v
\end{array}\right)
\left(
\begin{array}{c}\chi_1\\ \chi_2
\end{array}\right)
\eeq
and
\beq
\left(
\begin{array}{c}G^\pm\\ H^\pm
\end{array}\right)
=
\left(
\begin{array}{cc} v_1/v & \quad v_2/v\\ -v_2/v &\quad  v_1/v
\end{array}\right)
\left(
\begin{array}{c}\varphi_1^\pm\\ \varphi_2^\pm
\end{array}\right)
\eeq
in order to extract $G^0$ and $G^\pm$ as the massless Goldstone fields, whereas $H^\pm$ are the massive charged scalars.

The model also contains three neutral scalars, which are linear combinations of the $\eta_i$,
\begin{equation} \label{Eq:R-def}
\begin{pmatrix}
H_1 \\ H_2 \\ H_3
\end{pmatrix}
=R
\begin{pmatrix}
\eta_1 \\ \eta_2 \\ \eta_3
\end{pmatrix},
\end{equation}
with the $3\times3$ orthogonal rotation matrix $R$ satisfying
\begin{equation}
\label{Eq:cal-M}
R{\cal M}^2R^{\rm T}={\cal M}^2_{\rm diag}={\rm diag}(M_1^2,M_2^2,M_3^2),
\end{equation}
and with $M_1\leq M_2\leq M_3$. The rotation matrix $R$ can conveniently be parametrized as \cite{Accomando:2006ga, ElKaffas:2006gdt}
\begin{equation} 
R=
\begin{pmatrix}
R_{11}    &  R_{12}   & R_{13}   \\
R_{21}    &  R_{22}   & R_{23}   \\
R_{31}    &  R_{32}   & R_{33}  
\end{pmatrix}
=
\begin{pmatrix} \label{Rmatrix}
c_1\,c_2 & s_1\,c_2 & s_2 \\
- (c_1\,s_2\,s_3 + s_1\,c_3) 
& c_1\,c_3 - s_1\,s_2\,s_3 & c_2\,s_3 \\
- c_1\,s_2\,c_3 + s_1\,s_3 
& - (c_1\,s_3 + s_1\,s_2\,c_3) & c_2\,c_3
\end{pmatrix}.
\end{equation}
Since $R$ is orthogonal, only three of the elements $R_{ij}$ are independent, the rest can be expressed by these through the use of orthogonality relations.
From the potential one can now derive expressions for the masses of the scalars as well as Feynman rules for  scalar interactions. For the general basis that we consider here,
these expressions are quite involved and lengthy, so for convenience we refer the reader to Appendix A of Ref.~\cite{Grzadkowski:2014ada} where they have been collected.

%%%%%%%%%%%%%%%%%%%%%%%%%%%%%%%%%%%%%%%%%%%%%%%%%%%%%%%%%%%%
\section{Input parameters}
\label{Sec:input}
\setcounter{equation}{0}
%%%%%%%%%%%%%%%%%%%%%%%%%%%%%%%%%%%%%%%%%%%%%%%%%%%%%%%%%%%%
For the input parameters of the 2HDM67 potential, following Ref.~\cite{Grzadkowski:2014ada}, we adopt 
\begin{equation} \label{Eq:p67}
{\cal P}_{67}\equiv\{M_{H^\pm}^2,\mu^2,M_1^2,M_2^2,M_3^2,{\rm Im}\lambda_5,{\rm Re}\lambda_6,{\rm Re}\lambda_7, v_1, v_2, \xi,\alpha_1,\alpha_2,\alpha_3\},
\end{equation}
a set of 14 independent parameters
where $\mu^2$ represents the real part of the bilinear mixing term $m_{12}^2$, $\xi$ is the relative phase of the VEVs $v_1$ and $v_2$, and the $\alpha_i$ parametrize the neutral-sector orthogonal rotation matrix $R$.
All other potential parameters could be calculated using the set ${\cal P}_{67}$, see appendix A in Ref.~\cite{Grzadkowski:2014ada}.

In the 2HDM these parameters will only appear in certain combinations, leaving us with a total of 11 observable physical quantities. These can be chosen to be the minimal set consisting of the four independent masses of the scalars along with seven independent couplings~\cite{Grzadkowski:2014ada,Grzadkowski:2016szj},
\begin{equation} \label{Eq:pcal}
{\cal P}\equiv\{M_{H^\pm}^2,M_1^2,M_2^2,M_3^2,e_1,e_2,e_3,q_1,q_2,q_3,q\},
\end{equation}
where $e_i\equiv v_1R_{i1}+v_2R_{i2}$ is a factor appearing in the $H_iW^+W^-$ coupling (and several other gauge couplings as well, see appendix \ref{sect:cubic_gauge_couplings}). They satisfy the relation $e_1^2+e_2^2+e_3^2=v^2$.
Furthermore, $q_i$ is the coefficient of the $H_iH^+H^-$ term in the potential and $q$ is the coefficient of the $H^+H^+H^-H^-$ term in the potential. Note that scalar masses and their couplings to vector bosons ($e_i$) are independent of each other. Nevertheless, as we will discuss below, they are subject to certain theoretical consistency constraints if perturbativity is supposed to hold.

There is an important comment in order here. It can be shown that the following useful relation holds\footnote{\Eq{miei} is equivalent to \eq{weightedsum}, which is expressed in terms of the scalar potential parameters defined in a basis where $v_1=v$, $v_2=0$ and $\xi=0$ (the so-called Higgs basis, which is treated in more detail in Appendix~\ref{HiggsBasis}).  Indeed, \eq{miei} reduces to \eq{weightedsum} upon making the substitutions, $\lambda_i=Z_i$, $v_1=v$, $v_2=0$ and $\xi=0$. \label{fnone}}
\bea 
e_1^2M_1^2+e_2^2M_2^2+e_3^2M_3^2&=&v_1^4\lambda_1+v_2^4\lambda_2
+2v_1^2v_2^2(\lambda_3+\lambda_4+\Re\left[e^{2i\xi}\lambda_5\right])\nonumber\\
&&+4v_1^3v_2
\Re\left[e^{i\xi}\lambda_6\right]+4v_1v_2^3
\Re\left[e^{i\xi}\lambda_7\right], \label{miei}
\eea
The above relation implies that if one requires the quartic coupling constants $\lambda_i$ to remain in a perturbative regime, e.g. $\lambda_i < 4\pi$, then in the decoupling limit\footnote{The decoupling limit of the 2HDM was also discussed in Refs.~\cite{Haber:1989xc} and \cite{Gunion:2002zf}.
} of $M_{2,3,H^\pm} \to \infty$ the SM is recovered as the low-energy effective theory only for  $e_2=e_3=0$.\footnote{Due to the relation $e_1^2+e_2^2+e_3^2=v^2$, this implies that $e_1=v$, so that $H_1$ couples in exactly the SM manner.} Note also that if we had chosen $e_2=e_3=0$ (AL) as our starting point, then any value of $M_{2,3,H^\pm}>M_1$ would be allowed, in particular relatively light $H_{2,3}$ with $M_{2,3,H^\pm} \sim v$ would be a viable option.

For completeness, we note the following useful relations,\footnote{\Eqs{mchiggs}{m2sum} have been obtained from  eqs.~(A.4)--(A.7) of Ref.~\cite{Grzadkowski:2014ada} after employing the scalar potential minimum conditions. We also note that \eq{mchiggs} is equivalent to eqs.~(2.17) and (2.21) of Ref.~\cite{Pilaftsis:1999qt}.}
\bea
M_{H^\pm}^2&=&\frac{v^2}{2v_1v_2}\Re\left[(m_{12}^2 -v_1^2\lambda_6-v_2^2\lambda_7)e^{i\xi}-v_1 v_2(\lambda_4+\lambda_5 e^{2i\xi})\right]\,,\label {mchiggs}\\
M_1^2+M_2^2+M_3^2&=& \frac{v^2}{v_1 v_2}\Re(m_{12}^2 e^{i\xi})+v_1^2\lambda_1+v_2^2\lambda_2-v^2\Re(\lambda_5 e^{2i\xi} )
\nonumber \\
&&\qquad
 -(v_1^2-v_2^2)\left[\frac{v_1}{v_2}\Re(\lambda_6 e^{i\xi})-\frac{v_2}{v_1}\Re(\lambda_7 e^{i\xi})\right]\,.\label{m2sum}
\eea
The expressions for \eqs{mchiggs}{m2sum} in terms of Higgs basis parameters are given in \eqs{mch}{mtrace} of Appendix~\ref{HiggsBasis}.  Note that it is not so easy to obtain these results directly from \eqs{mchiggs}{m2sum} as we did in footnote~\ref{fnone}.  However, by employing the scalar potential minimum conditions given in eqs.~(A1)--(A3) of Ref.~\cite{Grzadkowski:2014ada}, one can re-express $\Re(m_{12}^2 e^{i\xi})$ in terms of $m_{11}^2+m_{22}^2$ and the $\lambda_i$.   Employing this result in \eqs{mchiggs}{m2sum} yields their equivalent form,\footnote{Indeed \eqs{mhpm}{m1m2m3} reduce to \eqs{mchiggs}{m2sum}, respectively, after making the substitutions, $m_{11}^2=-2Y_1$, $m_{22}^2=-2Y_2$, $\lambda_i=Z_i$, $v_1=v$, $v_2=0$, $\xi=0$ and employing the scalar potential minimum condition given in \eq{yz}.
}
\bea
M_{H^\pm}^2&=&\half\bigl(\lambda_1 v_1^2+\lambda_2 v_2^2+\lambda_3 v^2)+v_1 v_2\Re[(\lambda_6+\lambda_7)e^{i\xi}]-\half(m_{11}^2+m_{22}^2)\,, \label{mhpm}\\[6pt] 
M_1^2+M_2^2+M_3^2&=& 2(\lambda_1 v_1^2+\lambda_2 v_2^2)+v^2(\lambda_3+\lambda_4)+4v_1 v_2\Re[(\lambda_6+\lambda_7)e^{i\xi}]-m_{11}^2-m_{22}^2\,.\nonumber \\
&&\phantom{line}\label{m1m2m3}
\eea
The above equations will prove to be useful in the context of discussing alignment with or without decoupling in sec.~\ref{Sec:Bosonic_decays}.

The above shows that at fixed values of the $\lambda_i$, increasing values of $M_{2,3}$ and $M_{H^\pm}$ require positive and increasing $\Re m_{12}^2$.
Note however, that the SM would be recovered only if $e_2=e_3=0$ was chosen. 

Different bases for $(\Phi_1,\Phi_2)$ could be adopted while discussing the model, this freedom is 
parametrized by the following $U(2)$
transformation:
\beq
\left(
\begin{array}{c}\bar{\Phi}_1\\ \bar{\Phi}_2
\end{array}\right)
=
e^{i\psi}\left(
\begin{array}{cc}\cos\theta & e^{-i\tilde\xi}\sin\theta\\ -e^{i\chi}\sin\theta & e^{i(\chi-\tilde\xi)}\cos\theta
\end{array}\right)
\left(
\begin{array}{c}\Phi_1\\ \Phi_2
\end{array}\right)\equiv
U\left(
\begin{array}{c}\Phi_1\\ \Phi_2
\end{array}\right).
\label{U(2)}
\eeq

All the parameters of $\pcal$ are invariant under a change of basis, hence they represent observables of the 2HDM. 
Note that apart from the overall phase $\psi$, the $U$ matrix has 3 parameters, matching the reduction from the 14 potential parameters to the 11 physical parameters of ${\cal P}$. In Appendices \ref{sect:cubic_couplings} and \ref{sect:quartic_couplings} we see that we can express all the real couplings in terms of the parameters of $\pcal$, meaning that all the real couplings in the 2HDM represent observables of the model. In Appendix~\ref{HiggsBasis} we elucidate the connection to the Higgs basis \cite{Donoghue:1978cj,Georgi:1978ri,BLS,Davidson:2005cw}.

There are also complex couplings (both scalar and gauge) in which we need the auxiliary quantities $f_j\equiv v_1R_{j2}-v_2R_{j1}-ivR_{j3}$ \cite{Grzadkowski:2014ada} in the expressions. These are not basis invariant, they are what is referred to as pseudo-invariants under a change of basis. That means that they acquire a phase factor under a change of basis, i.e.
\bea
f_j\xrightarrow{\text{Basis change}}\bar{f}_j=f_j e^{i\delta}.
\eea
The phase $\delta$ depends on the $U(2)$-transformation we use to change basis, but is independent of $j$, meaning that all three $f_j$ acquire the same phase factor under a change of basis. An explicit expression for $e^{i\delta}$ is given in eq.~(\ref{expidelta}), in terms of the $U$-matrix, and the phases $\xi_j$ of eq.~(\ref{vevs}).

Since the $f_j$ are not invariant under a basis change, they do not represent observables of the theory. However, we may combine pseudo-invariants into something that is invariant by pairing it with one of its complex conjugate partners, i.e. 
\bea
f_if_j^*&=&v^2\delta_{ij}-e_ie_j+iv\epsilon_{ijk}e_k.
\label{fifj}
\eea
The combination $f_if_j^*$ is obviously basis invariant, and we see explicitly that it can be expressed in terms of the parameters of $\pcal$. 
It is also clear that the absolute values $|f_i|$ are physical (since they are basis independent) and also, as seen from eq.~(\ref{fifj}), could be expressed through other 
parameters already present in ${\cal P}$. This is consistent with the fact that the model has only 11 physical parameters originating from the potential.

As will be shown in appendix~\ref{fj_trans} the phase $\delta$ could be totally removed from the Lagrangian by a rephasing of the charged scalar field $H^\pm$, so that in effect $f_i$ could be considered as ``invariant" under a basis transformation that is accompanied by a rephasing of $H^\pm$.

A relevant question to ask is whether constraints we put on our set of parameters merely amount to choosing a basis, or whether they are in fact constraints on the model itself. If we put constraints on the parameters of ${\cal P}_{67}$ in such a way that all eleven parameters of ${\cal P}$ are still free to choose independently, then our constraints merely amount to a choice of basis\footnote{A popular choice of basis is the Higgs basis, which is discussed in more detail in Appendix~\ref{HiggsBasis}. In this basis only the first doublet has a VEV, meaning that $\xi=0$ and $v_2=0$, implying $v_1=v$. There is still some freedom left in performing a $U(1)$-rotation on $\Phi_2$. This can for instance be used to make $m_{12}^2$ real. All the eleven parameters of ${\cal P}$ are still free and independent of each other, so we have in choosing the Higgs basis in no way constrained the model.}. If on the other hand our constraints in some way limit the 11 parameters of ${\cal P}$ in such a way that they are not all free to choose independently anymore, then we have in fact constrained the model.\footnote{As we shall soon see, exact alignment is equivalent to putting $\alpha_1=\arctan(v_2/v_1)=\beta$ and $\alpha_2=0$. This in turn implies $e_1=v$ and $e_2=e_3=0$. Thus, alignment fixes some of the physical observables of ${\cal P}$, and therefore represents a constraint on the model as opposed to a choice of basis.}

%%%%%%%%%%%%%%%%%%%%%%%%%%%%%%%%%%%%%%%%%%%%%%%%%%%%%%%%%%%%
\section{The alignment limit}
\label{Sec:alignment}
\setcounter{equation}{0}
%%%%%%%%%%%%%%%%%%%%%%%%%%%%%%%%%%%%%%%%%%%%%%%%%%%%%%%%%%%%
The coupling between the lightest neutral Higgs boson $H_1$ and vector bosons is parametrized by \cite{Grzadkowski:2014ada}
\beq
e_1 = v \cos(\alpha_2)\cos(\alpha_1-\beta) ,
\label{Eq:e_1}
\eeq
where $\tan\beta=v_2/v_1$. As we have already stated, alignment is equivalent to $e_1=v$, $e_2=e_3=0$, when expressed in terms of the parameter set $\pcal$, implying 
\beq \label{Eq:h1sm-limit}
\alpha_1=\beta, \quad \alpha_2=0.
\eeq
The rotation matrix in this case becomes
\begin{equation} 
R=
\begin{pmatrix}
R_{11}    &  R_{12}   & R_{13}   \\
R_{21}    &  R_{22}   & R_{23}   \\
R_{31}    &  R_{32}   & R_{33}  
\end{pmatrix}
=
\begin{pmatrix}
c_\beta & s_\beta & 0 \\
-  s_\beta\,c_3 & c_\beta\,c_3  & s_3 \\
s_\beta\,s_3 & - c_\beta\,s_3  & c_3
\end{pmatrix}.
\label{R_lim}
\end{equation}
So that the mixing matrix could be written as  
\begin{equation} 
R= R_3 R_1 =
\begin{pmatrix}
1 & 0 & 0 \\
0 & c_3  & s_3 \\
0 & -s_3 & c_3
\end{pmatrix}
\begin{pmatrix}
c_\beta & s_\beta & 0 \\
-  s_\beta\ & c_\beta  & 0 \\
0 & 0 & 1
\end{pmatrix}.
\end{equation}

\noi Furthermore, in the AL we have
\begin{equation}
	f_1=0,\quad f_2=if_3\equiv\tilde{f}=v(c_3-is_3)=ve^{-i\alpha_3}.
\end{equation}

Later on in this paper the Type I and II versions of the 2HDM will be considered as reference models\footnote{When referring to
the scalar potential of those models we will either be using the term ``model with softly broken $\zBB_2$ symmetry''  or ``2HDM5''.}, therefore it is useful to recall here constraints
that emerge as consequences of the $\zBB_2$ symmetry imposed on the dimension-4 part of the potential. Then $\lambda_6=\lambda_7=0$
and consequently the $(1,3)$ and $(2,3)$ entries of the neutral mass-squared matrix, ${\cal M}^2_{13}$ and ${\cal M}^2_{23}$, are related as follows
\begin{equation} 
{\cal M}^2_{13} = t_\beta {\cal M}^2_{23},
\label{corel}
\end{equation}
where $t_\beta\equiv \tan\beta$. 
As a consequence of the above relation there is a constraint that relates mass eigenvalues, mixing angles and $t_\beta$~\cite{Khater:2003wq}:
\begin{equation} 
M_1^2 R_{13}(R_{12}t_\beta-R_{11}) + M_2^2 R_{23}(R_{22} t_\beta-R_{21}) + M_3^2 R_{33}(R_{32}t_\beta-R_{31}) = 0.
\end{equation}
In the AL, the above relation simplifies to
\beq
(M_2^2-M_3^2) s_3 c_3 s_\beta = 0,
\label{magic}
\eeq
so that either $M_2=M_3$, $\alpha_3=0$ or $\alpha_3=\pm\pi/2$. Here, we assume no mass degeneracy, $M_2\neq M_3$, so $\alpha_3=0$ or $\alpha_3=\pm\pi/2$. As will be discussed below, the two possible choices of $\alpha_3$ correspond to two possible CP-conserving versions of the 2HDM5 with different neutral-boson mass orderings.  

%%%%%%%%%%%%%%%%%%%%%%%%%%%%%%%%%%%%%%%%%%%%%%%%%%%%%%%%%%%%
\subsection{Alignment with or without decoupling}
\label{ALDL}
%%%%%%%%%%%%%%%%%%%%%%%%%%%%%%%%%%%%%%%%%%%%%%%%%%%%%%%%%%%%

We previously remarked that the exact alignment limit, where $e_2=e_3=0$, is realized in the decoupling limit of $M_{2,3,H^\pm}\to\infty$ where the quartic coupling constants $\lambda_i$ are held fixed.  More precisely, if
$M_{2,3,H^\pm}\gg v$, it follows that $|e_2/v|$, $|e_3/v|\ll 1$, which implies that the tree-level properties of $H_1$ are SM-like.  Thus, in the decoupling regime, the alignment limit is approximately realized.  

Nevertheless, there is a physical distinction between the alignment limit in the decoupling regime and the alignment limit without decoupling.  In either case, one must have $|e_2/v|$, $|e_3/v|\ll 1$, which means that the distinction between alignment with or without decoupling cannot be detected via the tree-level Higgs couplings to gauge bosons and fermions.   However, the distinction is present in the cubic and quartic tree-level Higgs couplings.   This is most clearly illustrated by examining the cubic $H_1 H_1 H_1$ coupling in the alignment limit.  Starting from the exact expression given in \eq{hhh1},
one finds that in the approximate alignment limit, 
\begin{equation} \label{h1h1h1}
H_1 H_1 H_1: \quad \frac{M_1^2}{2v}-\frac{(e_2^2+e_3^2)M^2_{H^\pm}}{v^3}\,.
\end{equation}
In the limit of alignment without decoupling, $M_{H^\pm}^2/v^2\sim\mathcal{O}(1)$, in which case, the correction to the exact AL result of $M_1^2/(2v)$ is quadratic in the small parameters $e_2/v$, $e_3/v$.  In contrast, in the limit of alignment with decoupling, 
\begin{equation}
e_2 M_{H^\pm}^2/v \sim\mathcal{O}(1)\,,\qquad\quad  e_3 M_{H^\pm}^2/v\sim\mathcal{O}(1)\,,
\end{equation}
as shown explicitly in \eq{decouplim}.  In this case, in \eq{h1h1h1} the correction to the exact AL result of $M_1^2/(2v)$ is linear in the small parameters $e_2/v$, $e_3/v$.

One additional distinction between alignment with or without decoupling arises when radiative corrections are taken into account.   In the limit of alignment without decoupling, the effects of loops containing $H_2$, $H_3$ and $H^\pm$ can compete with electroweak loop effects.  For example, the decay width of $H_1\to\gamma\gamma$ in the alignment limit can deviate from its SM value due to the effects of a charged Higgs boson loop \cite{Bhattacharyya:2014oka,Bernon:2015qea}.   In contrast, in the decoupling limit, the effects of heavy Higgs contributions in loop diagrams decouple.  Thus, in the previously cited example of $H_1\to\gamma\gamma$, the corresponding decay width approaches its SM value in the decoupling limit. 

%%%%%%%%%%%%%%%%%%%%%%%%%%%%%%%%%%%%%%%%%%%%%%%%%%%%%%%%%%%%
\subsection{Scalar couplings}
\label{Sec:scal_couplings}
%%%%%%%%%%%%%%%%%%%%%%%%%%%%%%%%%%%%%%%%%%%%%%%%%%%%%%%%%%%%
In the AL the scalar, $H_iH^+H^-$, couplings $q_i$ could be expressed through the mixing angle $\alpha_3$ and other parameters as follows~\cite{Grzadkowski:2014ada}\footnote{Here we adopt a weak basis such that the relative phase of the two VEVs vanishes, i.e. $\xi=0$.}
\begin{eqnarray}
q_{1}
&=&\frac{1}{v}\left(2 M_{H^\pm}^2-2\mu^2+M_1^2\right),
\label{q1}\\
q_{2}
&=&
+c_3\left[\frac{(\cb^2-\sb^2)}{v \cb \sb}(M_2^2-\mu^2)
+\frac{v}{2 \sb^2 }\Re\lambda_6
-\frac{v}{2 \cb^2}\Re\lambda_7\right]
+s_3\frac{v}{2 \cb \sb}\Im\lambda_5,
\label{q2}\\
q_{3}
&=&
-s_3\left[\frac{(\cb^2-\sb^2)}{v \cb \sb}(M_3^2-\mu^2)
+\frac{v}{2 \sb^2}\Re\lambda_6
-\frac{v}{2 \cb^2}\Re\lambda_7\right]
+c_3\frac{v}{2 \cb \sb}\Im\lambda_5.
\label{q3}
\end{eqnarray}
As has been shown in Ref.~\cite{Grzadkowski:2014ada}, in the AL, CP violation may remain only in the
weak-basis invariant  $\Im J_{30}$:
\beq
\Im J_1 = 0, \lsp \Im J_2 = 0, \lsp \Im J_{30} = \frac{q_2 q_3}{v^4}(M_3^2-M_2^2).
\label{cpv_AL}
\eeq
Therefore we can conclude that if CP is conserved in the bosonic sector in the AL, then it forces $q_2 q_3$ to vanish. Remembering that $e_2=e_3=0$ in the AL, we may conclude that  either $H_2$ is CP-odd ($e_2=q_2=0$) and $H_3$ is CP-even, or vice versa (see  Ref.~\cite{Grzadkowski:2014ada}). Note that this is consistent with the presence of the $H_2H_3Z_\mu$ coupling (proportional to $e_1$).

In order for $H_2$ to be CP-odd we require $e_2=q_2=0$, and for $H_3$ to be CP-odd we require $e_3=q_3=0$. Since $e_2=e_3=0$  in the AL, 
it follows from \eq{q2} that $H_2$ is CP-odd in the AL if
\beq
\tan\alpha_3=-\frac{2\sb\cb(\cb^2-\sb^2)(M_2^2-\mu^2)
	+v^2(\cb^2\Re\lambda_6
	-\sb^2\Re\lambda_7)}{v^2\sb\cb\Im\lambda_5},
\eeq
assuming that the numerator and denominator above are not both zero.   In the special case just cited, $H_2$ is CP-odd in the AL if
\beq
\Im\lambda_5=2(c_\beta^2-s_\beta^2)(M_2^2-\mu^2)+\frac{v^2c_\beta}{s_\beta}\Re\lambda_6-\frac{v^2s_\beta}{c_\beta}\Re\lambda_7=0\,,
\eeq
independently of the value of $\alpha_3$.
Likewise, it follows from \eq{q3} that $H_3$ is CP-odd in the AL if
\beq
\tan\alpha_3=\frac{v^2\sb\cb\Im\lambda_5}{2\sb\cb(\cb^2-\sb^2)(M_3^2-\mu^2)
	+v^2(\cb^2\Re\lambda_6
	-\sb^2\Re\lambda_7)}\,,
\eeq
assuming that the numerator and denominator above are not both zero.   In the special case just cited, $H_3$ is CP-odd in the AL if
\beq
\Im\lambda_5=2(c_\beta^2-s_\beta^2)(M_3^2-\mu^2)+\frac{v^2c_\beta}{s_\beta}\Re\lambda_6-\frac{v^2s_\beta}{c_\beta}\Re\lambda_7=0\,,
\eeq
independently of the value of $\alpha_3$.
In particular, apart from the special cases noted above, we see that for a model in which $\Im\lambda_5=0$ in the AL,
\bea
H_2\text{ CP-odd: }&&\alpha_3=\pm\half\pi,\\
H_3\text{ CP-odd: }&&\alpha_3=0.
\eea
Note that in the generic 2HDM67, when $\alpha_3=0$ or $\pm\half\pi$ in the AL, the mixing matrix $R$ is block-diagonal (modulo basis reordering), parametrized by the angle $\beta$ only. Nevertheless at those parameter points CP is violated since $q_2q_3$ would in general be non-zero (unless additional conditions specified in sec.~\ref{com_AL} are satisfied).

In models with softly broken $\zBB_2$ symmetry one finds, adopting eq.~(3.6) of Ref.~\cite{ElKaffas:2007rq}, 
that when $\alpha_3=0$ or $\alpha_3=\pm\pi/2$ (as implied by eq.~(\ref{magic})) then
$\Im\lambda_5=0$. Therefore it is easy to see from eqs.~(\ref{q2})--(\ref{q3}) that in 
these cases $q_3=0$ or $q_2=0$, respectively. Note that when $q_3=0$ (and $q_2\neq 0$), since also $e_3=0$, then $H_3$ is CP-odd and $H_{1,2}$ are CP-even. Conversely, when $q_2=0$ (and $q_3\neq 0$), since also $e_2=0$, then $H_2$ is CP-odd and $H_{1,3}$ are CP-even. In other words, these limits reproduce two possible versions of the 2HDM5 with a pseudoscalar that
is the heaviest ($A=H_3$) or next to the heaviest ($A=H_2$). 

In Appendices \ref{sect:cubic_couplings} and \ref{sect:quartic_couplings}, we have expressed all the scalar couplings in terms of the eleven parameters of the minimal set $\pcal$ (and in addition the auxiliary quantities $f_i$). Here, we specialize these to the exact AL without decoupling, by simply using $e_1=v$, $e_2=e_3=0$.
For the purpose of presenting couplings in a compact way, we use the notation $i=1,2,3$, 
whereas $j$ and $k$ refer to either 2 or 3, but not to 1.  (In vertices that involve both $H_j$ and $H_k$, the couplings presented below also apply to the cases of $j=k=2,3$.)
The non-zero trilinear couplings become (couplings involving Goldstone bosons are not listed):
\begin{subequations}
\begin{alignat}{2}
&H_1H_1H_1:&\quad&\frac{M_1^2}{2v},\\
&H_jH_kH_k:&\quad&\frac{q_j}{2}, \label{h322}\\
&H_1H_jH_j:&\quad&\frac{q_1}{2}+\frac{M_j^2-M_{H^\pm}^2}{v},\\
&H_iH^+H^-:&\quad&q_i,
\label{h3pm}
\end{alignat}
\end{subequations}
whereas the corresponding non-vanishing quartic ones are
\begin{subequations}
\begin{alignat}{2}
&H_1H_1H_1H_1:&\quad&\frac{M_1^2}{8v^2},\\
&H_jH_jH_jH_j:&\quad&\frac{q}{4},\\
&H_1H_jH_kH_k:&\quad&\frac{q_j}{2v},\\
&H_1H_1H_jH_j:&\quad&\frac{q_1}{4v}+\frac{M_j^2-M_{H^\pm}^2}{2v^2},\\
&H_2 H_2 H_3 H_3:&\quad&\frac{q}{2},\\
&H_1H_1H^+H^-:&\quad&\frac{q_1}{2v},	\\
&H_jH_jH^+H^-:&\quad&q,	\\
&H_1H_jH^+H^-:&\quad&\frac{q_j}{v},\\
&H^+H^+H^-H^-:&\quad& q.
\end{alignat}
\end{subequations}
Note that if CP is conserved, so that $q_2q_3=0$, only those cubic and quartic couplings survive which are invariant with respect to CP, assuming that $H_2$ and $H_3$ have opposite CP parities.

If the alignment limit is realized in the decoupling regime, then one must allow for the possibility of contributions
of the form $e_2 M^2$ and $e_3 M^2$ (where $M=M_2$, $M_3$ or $M_{H^\pm}$), which do not vanish but approach a constant value as $e_{2,3}\to 0$ and $M\to\infty$.  This leads to the following additional non-zero trilinear and quadrilinear couplings,
\bea
H_1H_1H_k:&\quad&\frac{3e_k M_k^2}{2v^2},\label{h3pmDL} \\
H_1H_1H_1 H_k:&\quad&\frac{e_k M_k^2}{2v^3}\,.\label{h4pmDL}
\eea

%%%%%%%%%%%%%%%%%%%%%%%%%%%%%%%%%%%%%%%%%%%%%%%%%%%%%%%%%%%%
\subsection{Gauge couplings}
\label{Sec:vec_scal_couplings}
%%%%%%%%%%%%%%%%%%%%%%%%%%%%%%%%%%%%%%%%%%%%%%%%%%%%%%%%%%%%
Again, simply using the fact that in the AL $e_1=v$ and $e_2=e_3=0$, only the following gauge couplings
remain non-zero
(some vertices with corresponding Goldstone bosons are not shown)
\beq
H_1 Z_\mu Z^\mu:\quad 
\frac{g^2v}{4\cos^2\thetaW}, \quad H_1 W^+_\mu W^{-\mu}:\quad \frac{g^2v}{2}, 
\label{eq:H_iZZ}
\eeq
and
\begin{equation} \label{eq:e_i-epsilon}
(H_2\ddel_\mu \!H_3 )Z^\mu: \quad 
-\frac{g}{2\cos\thetaW}.
\end{equation}
Also,
\begin{subequations} \label{Eq:gaugecoulings}
\begin{alignat}{2}
H_2H^+A_\mu W^{-\mu}: \quad &
\frac{g^2}{2v}\sin\thetaW f_2, &\quad
H_2H^-A_\mu W^{+\mu}: \quad &
\frac{g^2v}{2}\sin\thetaW f_2^\ast, \\
H_2 H^+ Z_\mu W^{-\mu}: \quad &
\frac{-g^2}{2v}\frac{\sin^2\thetaW}{\cos\thetaW} f_2, &\quad
H_2 H^- Z_\mu W^{+\mu}: \quad &
\frac{-g^2}{2v}\frac{\sin^2\thetaW}{\cos\thetaW} f_2^\ast, \\
(H^+\ddel_\mu \!H_2 )W^{-\mu}: \quad &
i\frac{g}{2v}f_2, &\quad
(H^-\ddel_\mu\!H_2 )W^{+\mu}: \quad &
-i\frac{g}{2v}f^*_2.
\end{alignat}
\end{subequations}
with (in the AL) $f_2=ve^{-i\alpha_3}$.
In the AL, the corresponding couplings for $H_1$ do not exist, and those involving $H_3$ receive an extra phase factor $(f_3=-if_2)$.

Note that the $H_1 Z Z$, $H_1 W^+ W^-$ and $H_2H_3Z$ couplings of eqs.~(\ref{eq:H_iZZ})--(\ref{eq:e_i-epsilon}) do not involve $f_i$ and are CP-symmetric, assuming opposite parities for $H_2$ and $H_3$. The remaining scalar-gauge couplings of eq.~(\ref{Eq:gaugecoulings}) turn out to be invariant as well, however the CP transformation of the charged scalar field $H^+$ requires an extra phase $H^+ \stackrel{CP}{\rightarrow} e^{i\gamma}H^-$. Choosing e.g. $\gamma=2\alpha_3$ one finds that $H_2$ must be even and $H_3$ odd while for $\gamma=2\alpha_3+\pi$, the CP parities of $H_2$ and $H_3$ are reversed. The same could be concluded from another perspective. As we have already mentioned, the phase of $f_i$ depends on the weak basis, it turns out that it is possible to choose a basis such that e.g.\ this phase vanishes. In this particular basis the interactions of eq.~(\ref{Eq:gaugecoulings}) are symmetric under a standard  transformation of the charged scalar field $H^+ \stackrel{CP}{\rightarrow} H^-$. Both pictures are consistent with the general statement that the kinetic terms are CP-invariant.

%%%%%%%%%%%%%%%%%%%%%%%%%%%%%%%%%%%%%%%%%%%%%%%%%%%%%%%%%%%%
\subsection{General Yukawa couplings in the alignment limit}
\label{Sec:align_yukawa_couplings}
%%%%%%%%%%%%%%%%%%%%%%%%%%%%%%%%%%%%%%%%%%%%%%%%%%%%%%%%%%%%

Appendix~\ref{Yukawa-more} contains both the most general Yukawa couplings as well as various special cases. In this appendix we have parametrized the Yukawa matrices in terms of the two matrices $\kappa$, which simply becomes the diagonalized fermionic mass matrix, and the matrix $\rho$ which in the general case will be an arbitrary complex matrix. Special cases considered in the appendix include 
$\rho$-diagonal, Type I and Type II model Yukawa couplings. Here we focus on the AL couplings, so
$e_1=v$, $e_2=e_3=0$, and $f_1=0$, $f_2=ve^{-i\alpha_3}$, $f_3=-ive^{-i\alpha_3}$. This simplifies all Yukawa couplings to the neutral physical scalars.
In the AL, the phase factor $e^{-i\alpha_3}$ appears repeatedly in Yukawa couplings together with $\tilde{\rho}^f$, defined by eq.~(\ref{tilderho}).  Therefore it is convenient to define a related quantity $\bar\rho^f$ that absorbs the phase factor,
\begin{equation}
\bar\rho^f\equiv e^{-i\alpha_3}\tilde{\rho}^f,
\label{tilde}
\end{equation}
where $f=u,d,l$.

Specializing the generic results contained in appendix~\ref{Yuk_gen} one can write the couplings of the neutral Higgs boson in the AL as follows:
\begin{subequations} \label{yuk_ali}
\bea
\bar{f}_kf_mH_1:& &-\frac{m_{f_k}}{v}\delta_{km} \lsp ({\rm no~summation~over~}k),\\
\bar{l}_kl_mH_2:& &-\frac{1}{2\sqrt{2}}\left[\left(\bar\rho^{l\;*}_{mk}+\bar\rho^l_{km}\right)+\left(\bar\rho^{l\;*}_{mk}-\bar\rho^l_{km}\right)\gamma_5\right],\\
\bar{d}_kd_mH_2:& &-\frac{1}{2\sqrt{2}}\left[\left(\bar\rho^{d\;*}_{mk}+\bar\rho^d_{km}\right)+\left(\bar\rho^{d\;*}_{mk}-\bar\rho^d_{km}\right)\gamma_5\right],\\
\bar{u}_ku_mH_2:& &-\frac{1}{2\sqrt{2}}\left[\left(\bar\rho^{u\;*}_{mk}+\bar\rho^u_{km}\right)-\left(\bar\rho^{u\;*}_{mk}-\bar\rho^u_{km}\right)\gamma_5\right],\\
\bar{l}_kl_mH_3:& &-\frac{i}{2\sqrt{2}}\left[\left(\bar\rho^{l\;*}_{mk}-\bar\rho^l_{km}\right)+\left(\bar\rho^{l\;*}_{mk}+\bar\rho^l_{km}\right)\gamma_5\right],\\
\bar{d}_kd_mH_3:& &-\frac{i}{2\sqrt{2}}\left[\left(\bar\rho^{d\;*}_{mk}-\bar\rho^d_{km}\right)+\left(\bar\rho^{d\;*}_{mk}+\bar\rho^d_{km}\right)\gamma_5\right],\\
\bar{u}_ku_mH_3:& &-\frac{i}{2\sqrt{2}}\left[\left(\bar\rho^{u\;*}_{mk}-\bar\rho^u_{km}\right)-\left(\bar\rho^{u\;*}_{mk}+\bar\rho^u_{km}\right)\gamma_5\right].
\eea
\end{subequations}
Note that $H_1$ couples only flavor-diagonally in the AL, so indeed it behaves as a genuine SM Higgs boson. 

For the charged Higgs boson we obtain:
\begin{subequations}
\bea
\bar{\nu}_{k}l_mH^+:& &-\frac{1}{2}e^{-i\alpha_3}\bar\rho^{l\,*}_{mk}(1+\gamma_5),\\
\bar{l}_m\nu_{k}H^-:& &-\frac{1}{2} e^{i\alpha_3}\bar\rho^l_{mk}(1-\gamma_5),\\
\bar{u}_md_kH^+:& &\frac{1}{2}e^{-i\alpha_3}\left\{\left[(\bar\rho^u)^\dag K\right]_{mk}(1-\gamma_5)-\left[K(\bar\rho^d)^\dag\right]_{mk}(1+\gamma_5)\right\},\\
\bar{d}_ku_mH^-:& &\frac{1}{2} e^{i\alpha_3}\left\{\left[K^\dag\bar\rho^u\right]_{km}(1+\gamma_5)-\left[\bar\rho^d K^\dag\right]_{km}(1-\gamma_5)\right\}.
\eea
\label{char_yuk}
\end{subequations}
Note that the results contained  in eqs.~(\ref{yuk_ali})--(\ref{char_yuk}) are also applicable for the Type I and the Type II model by adopting the appropriate $\rho^f$ from appendices \ref{Yuk_Type_I} and \ref{Yuk_Type_II}, respectively, together with eqs.~(\ref{tilde}) and
(\ref{tilderho}).
A general (flavor non-diagonal) Yukawa coupling of the Higgs boson $H_\alpha$ could be written in the following form 
\begin{equation}
H_\alpha \bar f_k(a_{km}^{\alpha\, f}+i\gamma_5b_{km}^{\alpha\, f})f_m,
\label{yuk_gen}
\end{equation}
where $f=u,d,l$ with $\alpha=1,2,3$ and $a^{\alpha\, f}$ and $b^{\alpha\, f}$ hermitian matrices 
(as required by the hermiticity of the Yukawa Lagrangian) in the flavor space given by eqs.~(\ref{yuk_ali}).

Note that the following relations between scalar and pseudoscalar components of the $H_2$ and $H_3$ Yukawa couplings hold in the AL: 
\bea
a_{km}^{2\,l,d}&=\phantom{-}b_{km}^{3\,l,d}, \lsp a_{km}^{2\,u}&=-b_{km}^{3\,u},\\ 
b_{km}^{2\,l,d}&=-a_{km}^{3\,l,d}, \lsp b_{km}^{2\,u}&=\phantom{-}a_{km}^{3\,u}.
\label{yuk_rel}
\eea
Therefore the following sum rules are satisfied:\footnote{Similar sum rules applicable to a general $N$ Higgs doublet model have been presented in Ref.~\cite{Bento:2018fmy}.}
\bea
a_{km}^{2 \,f} a_{ij}^{2 \,f} + b_{km}^{2 \,f} b_{ij}^{2 \,f} &=& a_{km}^{3 \,f} a_{ij}^{3 \,f}+ b_{km}^{3 \,f} b_{ij}^{3 \,f},\label{sum_rule1} \\
a_{km}^{2 \,f} b_{ij}^{2 \,f} &=& - a_{km}^{3 \,f} b_{ij}^{3 \,f}. \label{sum_rule2}
\eea
The first sum rule is applicable for CP-conserving processes while the
second one is relevant for CP-violating observables. 
From eq.~(\ref{sum_rule2}) we can observe that in some sense the amount of CP violation (encoded by $a_{km}^{\alpha \,f} b_{ij}^{\alpha \,f}$) in the AL is opposite for $H_2$ and $H_3$.

It is worth discussing here CP properties of the general Yukawa couplings given by eq.~(\ref{yuk_gen}). Two cases must be considered with 
$H_\alpha$ being either even or odd under CP. Then CP conservation together with hermiticity of the Yukawa Lagrangian requires 
the following relations to hold:
\bea
&\!\!\!\!\!\!\!\!H_\alpha \overset{\rm CP}{\longrightarrow} +H_\alpha: a^{\alpha\,f}=\phantom{-}a^{\alpha\, f\, *}, \ssp a^{\alpha\,f}=\phantom{-}a^{\alpha\, f\, T}, \ssp
                                                              b^{\alpha\,f}=-b^{\alpha\, f\, *}, \ssp b^{\alpha\, f}=-b^{\alpha\, f\, T},&\\
&\!\!\!\!\!\!\!\!H_\alpha \overset{\rm CP}{\longrightarrow} -H_\alpha: a^{\alpha\,f}=-a^{\alpha\, f\, *}, \ssp a^{\alpha\,f}=-a^{\alpha\,f\, T}, \ssp
                                                              b^{\alpha\,f}=\phantom{-}b^{\alpha\, f\, *}, \ssp b^{\alpha\,f}=\phantom{-}b^{\alpha\,f\, T}.&
\eea
The above conditions could be expressed in terms of the $\bar\rho^{f}$ matrices.
Then, for instance for up-type quarks and $H_{2,3} \overset{\rm CP}{\longrightarrow} \pm H_{2,3}$, CP conservation of $H_2$ and $H_3$  Yukawa couplings would require
$\bar\rho^{u}=\pm\bar\rho^{u\;*}$ {\it and} $\bar\rho^{u}=\mp\bar\rho^{u\;*}$, respectively. Therefore if both $H_2$ {\it and} $H_3$ were CP-even or CP-odd\footnote{This case is considered just for completeness as in the CP-conserving limit of the 2HDM67, CP parities of $H_2$ and $H_3$ are indeed opposite with non-trivial Yukawa couplings.} then 
there would be no way to conserve CP in the Yukawa couplings of eq.~(\ref{yuk_gen}) unless $\bar\rho^{u} = 0$. However if $H_{2} \overset{\rm CP}{\longrightarrow} \pm H_{2}$ and $H_{3} \overset{\rm CP}{\longrightarrow} \mp H_{3}$ then CP is conserved in both couplings for $\bar\rho^{u}=\pm\bar\rho^{u\;*}$. Note that the CP-parities of $H_2$ and $H_3$ are fixed by the CP conservation for
a given type of fermions ($l$, $d$ or $u$), therefore for the remaining fermions there is no more sign freedom in relations between  $\bar\rho^{f}$ and
$\bar\rho^{f\;*}$, if the signs do not match, CP is violated. 

We have already observed in section~\ref{Sec:scal_couplings} that if CP was conserved in the bosonic sector then $H_2$ and $H_3$ would have opposite CP parity. Above, by considering Yukawa couplings, we have confirmed this observation. 
 
Next, in order to proceed with some semi-qualitative discussion (see section~\ref{sect:fermionic_decays}), we assume that the $\bar\rho^f_{km}$ are flavor-diagonal matrices parametrized by 
\begin{subequations} \label{simple_yuk}
\begin{alignat}{2}
\bar\rho^l_{kk}&=\sqrt{2}\,\frac{m_\tau}{v}\hat\rho^l \delta_{k3},\\
\bar\rho^d_{kk}&=\sqrt{2}\,\frac{m_b}{v}\hat\rho^d \delta_{k3}, \\
\bar\rho^u_{kk}&=\sqrt{2}\,\frac{m_t}{v}\hat\rho^u \delta_{k3},
\end{alignat}
\end{subequations}
where $\hat\rho^f=|\hat\rho^f|e^{i \theta_f}$ are complex numbers.
Note that, even though it is a quite radical assumption, this way the $u$, $d$ and $l$ components of the Yukawa couplings are still independent, in contrast to what is observed in e.g.\ Type I and Type II models, where all the couplings are determined by $\tan\beta$ and fermion masses, see eqs.~(\ref{rhou_I})--(\ref{rhol_I}). Since $\bar\rho_{33}^f$ are complex numbers therefore, within this assumption we obtain 6 real free parameters to specify all the Yukawa couplings. 

It is worth stressing that in the generic model the Yukawa couplings are, in general, not proportional to fermion masses any more. In the AL, as seen from eq.~(\ref{yuk_ali}), although $H_1$ couplings are still proportional to the corresponding fermion masses, however those of $H_2$ and $H_3$ are not related to fermion masses at all. Since $e_2=e_3=0$, they are just parametrized by elements of the 
$\bar\rho^{f}$ matrix. Our choice of non-zero entries of $\bar\rho^f$ in eq.~(\ref{simple_yuk}) for only the third generation is dictated just by the fact that in the familiar Type I and Type II models the third family contributions dominate and therefore this approach facilitates comparison with $\zBB_2$ symmetric versions of the 2HDM. Other entries of $\bar\rho^{f}$ are also allowed.

%%%%%%%%%%%%%%%%%%%%%%%%%%%%%%%%%%%%%%%%%%%%%%%%%%%%%%%%%%%%
\subsection{CP properties at \boldmath{$\alpha_3=0,\pm \pi/2$} in the alignment limit}
\label{com_AL}
%%%%%%%%%%%%%%%%%%%%%%%%%%%%%%%%%%%%%%%%%%%%%%%%%%%%%%%%%%%%

As has already been noted above, in the AL the points corresponding to $\alpha_3=0,\pm \pi/2$ deserve special attention, as at those points the rotation matrix $R$ turns out
to be block-diagonal (modulo basis reordering).
Therefore here, assuming exact alignment ($\alpha_1=\beta$ and $\alpha_2=0$), we summarize CP properties  
of the $\zBB_2$-non-symmetric model and of the one with the $\zBB_2$ softly broken:
\bit
\item 
Assume $\zBB_2$ is broken by dim~4 terms, e.g., by $\lambda_6,\lambda_7\neq 0$. The remaining mixing angle $\alpha_3$ varies in the interval $[-\pi/2,\pi/2]$ and in general CP is violated. Adopting eqs.~ (\ref{q2})--(\ref{q3}) it is easy to verify that the condition for CPC is not satisfied even if $\alpha_3=0,\pm \pi/2$. In order to ensure CPC at those points one would have to assume in addition that either $\Im\lambda_5=0$ or 
\bit
\item [$\star$] $\frac{(\cb^2-\sb^2)}{v \cb \sb}(M_2^2-\mu^2)
+\frac{v}{2 \sb^2 }\Re\lambda_6
-\frac{v}{2 \cb^2}\Re\lambda_7=0$ for $\alpha_3=0$, 
\item [$\star$] $\frac{(\cb^2-\sb^2)}{v \cb \sb}(M_3^2-\mu^2)
+\frac{v}{2 \sb^2}\Re\lambda_6
-\frac{v}{2 \cb^2}\Re\lambda_7=0$ for $\alpha_3=\pm\pi/2$.
\eit
\item 
If, on the other hand, the $\zBB_2$ is only broken by the dim~2 term $m_{12}^2$ (with $\lambda_6=\lambda_7 = 0$), 
one finds by virtue of eq.~(\ref{magic}) that $\alpha_3=0,\pm \pi/2$. It turns out 
then (see eq.~(3.6) in Ref.~\cite{ElKaffas:2007rq}) that in the AL $\Im\lambda_5=0$ and therefore from eqs.~(\ref{q2})--(\ref{q3})  one finds that $q_2q_3=0$, so that CP is conserved at those points. 
\eit
%%%%%%%%%%%%%%%%%%%%%%%%%%%%%%%%%%%%%%%%%%%%%%%%%%%%%%%%%%%%%
\section{``Heavy'' Higgs production}
\label{Sec:production}
\setcounter{equation}{0}
%%%%%%%%%%%%%%%%%%%%%%%%%%%%%%%%%%%%%%%%%%%%%%%%%%%%%%%%%%%%%
Experimental searches are based on the assumption of production dominantly via either of two channels:
\begin{itemize}
\item
glue-glue fusion
\begin{equation}
gg\to H_i
\end{equation}
via quark triangle diagrams. The production cross section for $H_2 $ in the AL is then proportional to 
\begin{equation}
X_2^2 = \half\Bigl|\sum_{q,k} \Re[\bar\rho_{kk}^q]\frac{v}{m_{q_k}}A(\tau_{q_k})\Bigr|^2
+\half\Bigl|\sum_{q,k} \Im[\bar\rho_{kk}^q]\frac{v}{m_{q_k}}B(\tau_{q_k})\Bigr|^2,
\label{Eq:cross-section}
\end{equation}
and similarly for $H_3$ production ($X_3^2$), after the substitution 
$\Re[\bar\rho_{kk}^q] \longrightarrow \Im[\bar\rho_{kk}^q]$ and
$\Im[\bar\rho_{kk}^q] \longrightarrow -\Re[\bar\rho_{kk}^q]$.
The functions $A$ and $B$ are defined in appendix~\ref{sect:A_B_functions}.
\item
weak-boson fusion
\begin{equation}
q_1 q_2 \to H_j q_1^\prime q_2^\prime.
\end{equation}
This mechanism relies on the $W^+W^-H_i$ or $ZZH_i$ coupling, and thus does not contribute to the $H_2$ and $H_3$ production in the AL.
\end{itemize}
%%%%%%%%%%%%%%%%%%%%%%%%%%%%%%%%%%%%%%%%%%%%%%%%%%%%%%%%%%%%%
\begin{figure}[h!]
\begin{center}
\includegraphics[scale=0.80]{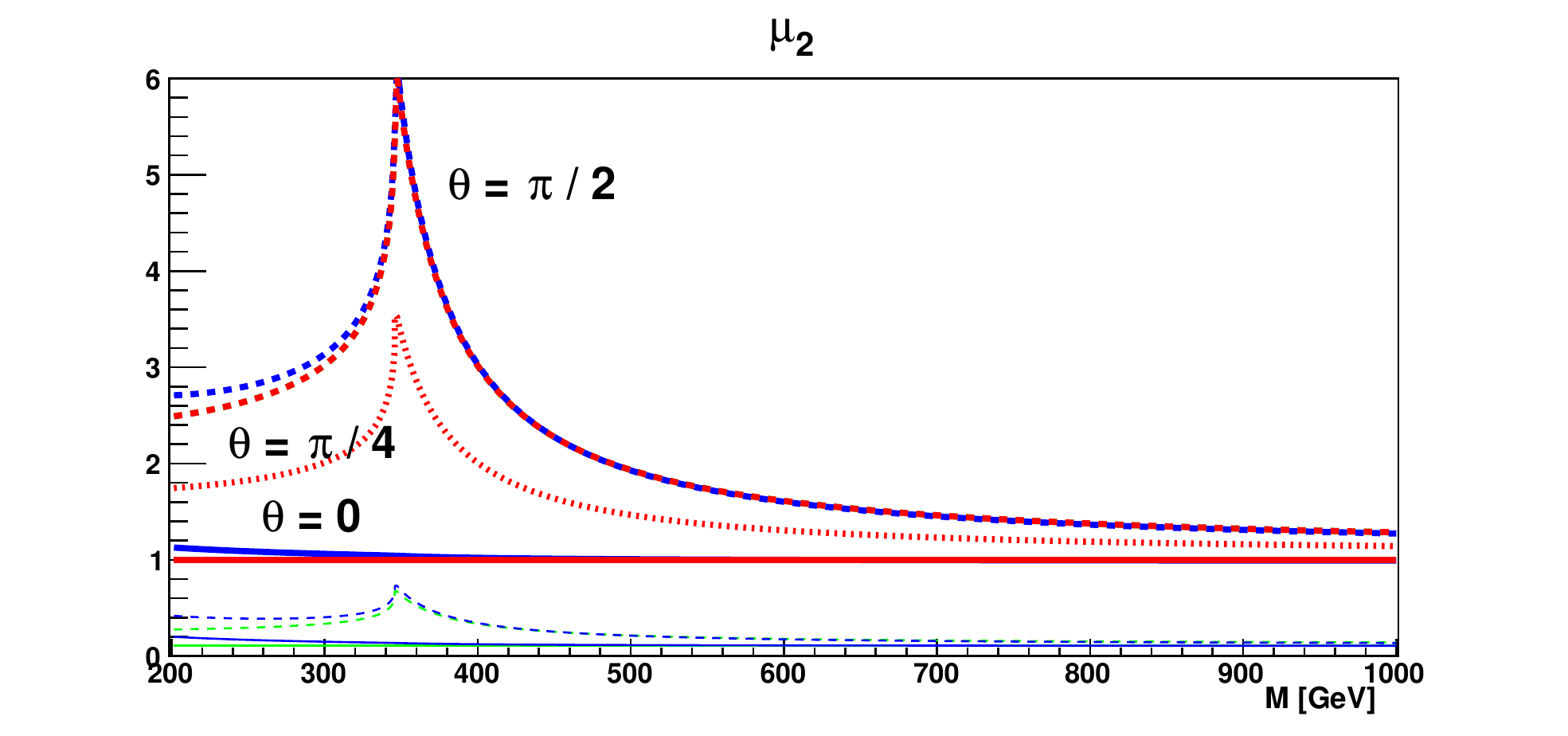}
\end{center}
\vspace*{-4mm}
\caption{Cross section ratios $\mu_2$ for $gg$ production of $H_2$ for $\bar\rho^q$ defined by  eq.~(\ref{simple_yuk}).
Here, for illustration we assume that phases of $\hat\rho^q$ are the same for up- and down-type quarks, 
so $\theta_u=\theta_d=\theta$.
Red, heavy: general model with $|\hat\rho^u|=|\hat\rho^d|=1$ with (solid) $\theta=0$,  (dotted) $\theta=\pi/4$ and (dashed) $\theta=\pi/2$.
Blue (green) heavy solid: Type~II (Type~I) for $\tan\beta=1$ with $\alpha_3 = 0$ (so $H_3 = A$). 
Dashed, same with $\alpha_3=\pi/2$ (so $H_2 = A$). Thin (blue and green) Type I and Type II with $\tan\beta=3$. (The green curves, for Type I, are partly covered by the red and blue ones.) The spike at $M=2m_t$ originates from the function $\Re B$ of eq.~(\ref{Eq:B}) describing the pseudoscalar coupling.}
\label{Fig:sigma}
\end{figure}
%%%%%%%%%%%%%%%%%%%%%%%%%%%%%%%%%%%%%%%%%%%%%%%%%%%%%%%%%%%%%

To illustrate effects of generic Yukawa couplings in the production of $H_{2,3}$ we define
\begin{equation}
\mu_{\alpha} \equiv \frac{\sigma_{\alpha}}{\sigma_\text{SM}}
=\frac{X_{\alpha}^2}{|A|^2_\text{SM}},
\end{equation}
where $\alpha=2,3$, $\sigma_\alpha$ is the $pp\to H_\alpha$ $gg$ cross section in the 2HDM, $\sigma_{\text{SM}}$ is the corresponding SM cross section, and $X_2^2$ is given by eq.~ (\ref{Eq:cross-section}). Here, $A_\text{SM}$ refers to the function (\ref{Eq:A}), summed over $t$ and $b$-quark loops.
In Fig.~\ref{Fig:sigma} we plot this quantity for $\alpha=2$.

It should be kept in mind that the normalization of this cross section depends critically on the assumed magnitudes $|\bar\rho|$, in particular on $|\bar\rho^u|$.
The assumption of eq.~(\ref{simple_yuk}) reproduces the predictions of the Type~I and Type~II models with $\tan\beta=1$, unless $\bar\rho$ has a non-zero phase $\theta$. In the latter case, if $\theta_t$ is non-zero, the production cross section would have a spike at the $t\bar t$ threshold.
%%%%%%%%%%%%%%%%%%%%%%%%%%%%%%%%%%%%%%%%%%%%%%%%%%%%%%%%%%%%%
\section{``Heavy'' Higgs boson decays in the approximate AL}
\label{Heavy_Higgs_decays}
\setcounter{equation}{0}
%%%%%%%%%%%%%%%%%%%%%%%%%%%%%%%%%%%%%%%%%%%%%%%%%%%%%%%%%%%%%

It is important to realize that in the AL the $H_2$ and $H_3$ couplings are strongly correlated allowing for construction of observables that may efficiently test the alignment scenario. 

There are four classes of interesting decay modes: final fermion-antifermion pairs, purely scalar decays, and final states involving a heavy gauge boson. We shall discuss them in the following subsections.

%%%%%%%%%%%%%%%%%%%%%%%%%%%%%%%%%%%%%%%%%%%%%%%%%%%%%%%%%%%%%
\subsection{Fermionic modes}
\label{sect:fermionic_decays}
%%%%%%%%%%%%%%%%%%%%%%%%%%%%%%%%%%%%%%%%%%%%%%%%%%%%%%%%%%%%

Table~\ref{tab_coupl} summarizes Yukawa couplings in the AL for the generic model adopting the assumption of eq.~(\ref{simple_yuk}) together 
with corresponding couplings for the Type I and Type II models (consistently also in the AL) used below as reference models to compare with 2HDM67 results.
Note the 2HDM67 Yukawa couplings for the case of eq.~(\ref{simple_yuk}) are parametrized by the quark masses and
six additional
independent numbers ($\Re\hat\rho^l$, $\Im\hat\rho^l$, $\Re\hat\rho^d$, $\Im\hat\rho^d$, $\Re\hat\rho^u$, $\Im\hat\rho^u$)
while for the Type I or Type II the freedom is much more limited as the couplings can be specified by only two parameters ($\beta$, $\alpha_3$) with $\alpha_3=0,\pm\pi/2$.

Adopting eq.~(\ref{simple_yuk}) one finds
\bea
\Gamma(H_2\to f_3\bar{f}_3)&=&\frac{3M_2}{8\pi}\frac{m_{f_3}^2}{v^2}\left(
(\Re \hat{\rho}^f)^2\beta_{f_3 2}^2+(\Im \hat{\rho}^f)^2
\right)\beta_{f_3 2},\\
\Gamma(H_3\to f_3\bar{f}_3)&=&\frac{3M_3}{8\pi}\frac{m_{f_3}^2}{v^2}\left(
(\Im \hat{\rho}^f)^2\beta_{f_3 3}^2+(\Re \hat{\rho}^f)^2
\right)\beta_{f_3 3}.
\eea
for $f_3=\tau,b,t$ and $\beta_{f\, \alpha}\equiv \sqrt{1-4m_f^2/M_\alpha^2}$.   With the exception of the case of
final state top quarks, we may approximate
$\beta_{f\, \alpha} \sim 1$. 
Note that when $\beta_{f\, \alpha} \sim 1$ then the relations of eq.~(\ref{yuk_rel}) imply that 
the squared matrix elements for $H_2 \to f \bar f$ and $H_3 \to f \bar f$ are nearly the same, 
so that the corresponding widths differ only by the overall scalar masses, therefore for $f_3\neq t$ one expects
\beq
\frac{\Gamma(H_2 \to f_3\bar f_3)}{\Gamma(H_3 \to f_3\bar f_3)} = \frac{M_2}{M_3} + {\cal{O}}\left(\frac{m_{f_3}^2}{M_{2,3}^2}\right).
\label{light_ratio} 
\eeq

\begin{table}[t]
\begin{center}
  \begin{tabular}{| c || c | c | c| }
    \hline
     & Type I & Type II & 2HDM67 \\ \hline\hline
    $a_{kk}^{2\,d}$ & $-\frac{m_{d_k}}{v}c_3\frac{c_\beta}{s_\beta}$           & $\phantom{-}\frac{m_{d_k}}{v}c_3\frac{s_\beta}{c_\beta}$ & $-\delta_{k3}\frac{m_b}{v}\Re\hat\rho^d$ \\ \hline
    $a_{kk}^{3\,d}$ & $\phantom{-}\frac{m_{d_k}}{v}s_3\frac{c_\beta}{s_\beta}$ & $-\frac{m_{d_k}}{v}s_3\frac{s_\beta}{c_\beta}$ & $-\delta_{k3}\frac{m_b}{v}\Im\hat\rho^d$ \\ \hline
    $b_{kk}^{2\,d}$ & $-\frac{m_{d_k}}{v}s_3\frac{c_\beta}{s_\beta}$           & $\phantom{-}\frac{m_{d_k}}{v}s_3\frac{s_\beta}{c_\beta}$ & $\phantom{-}\delta_{k3}\frac{m_b}{v}\Im\hat\rho^d$ \\ \hline
    $b_{kk}^{3\,d}$ & $-\frac{m_{d_k}}{v}c_3\frac{c_\beta}{s_\beta}$           & $\phantom{-}\frac{m_{d_k}}{v}c_3\frac{s_\beta}{c_\beta}$ & $-\delta_{k3}\frac{m_b}{v}\Re\hat\rho^d$ \\ \hline
    $a_{kk}^{2\,u}$ & $-\frac{m_{u_k}}{v}c_3\frac{c_\beta}{s_\beta}$           & $-\frac{m_{u_k}}{v}c_3\frac{c_\beta}{s_\beta}$ & $-\delta_{k3}\frac{m_t}{v}\Re\hat\rho^u$ \\ \hline
    $a_{kk}^{3\,u}$ & $\phantom{-}\frac{m_{u_k}}{v}s_3\frac{c_\beta}{s_\beta}$ & $\phantom{-}\frac{m_{u_k}}{v}s_3\frac{c_\beta}{s_\beta}$ & $-\delta_{k3}\frac{m_t}{v}\Im\hat\rho^u$ \\ \hline
    $b_{kk}^{2\,u}$ & $\phantom{-}\frac{m_{u_k}}{v}s_3\frac{c_\beta}{s_\beta}$ & $\phantom{-}\frac{m_{u_k}}{v}s_3\frac{c_\beta}{s_\beta}$ & $-\delta_{k3}\frac{m_t}{v}\Im\hat\rho^u$ \\ \hline
    $b_{kk}^{3\,u}$ & $\phantom{-}\frac{m_{u_k}}{v}c_3\frac{c_\beta}{s_\beta}$ & $\phantom{-}\frac{m_{u_k}}{v}c_3\frac{c_\beta}{s_\beta}$ & $\phantom{-}\delta_{k3}\frac{m_t}{v}\Re\hat\rho^u$\\ 
    \hline
  \end{tabular}
	\caption{Couplings $a_{kk}^{\alpha\,q}$ and $b_{kk}^{\alpha\,q}$ for the Type I, Type II models and for the generic 2HDM in the AL. For 2HDM5 one should consider the two cases: $\alpha_3=0$ and $\alpha_3=\pm\pi/2$. For the 2HDM67 the assumption of eq.~(\ref{simple_yuk}) was adopted.  
	\label{tab_coupl} }
\end{center}
\end{table}

It is useful to define the reduced width for fermionic two-body Higgs boson decays,
\beq
\bar\Gamma(H_\alpha \to f\bar f) \equiv \frac{8\pi}{3M_\alpha} \Gamma(H_\alpha \to f \bar f)\left(\frac{v}{m_f}\right)^2\beta_{f\,\alpha}^{-1},
\label{red_wid}
\eeq
with $\alpha=2,3$.
In Table~\ref{tab:reducedwidth} we collect predictions for $\tau^+\tau^-$, $b \bar b$, $c \bar c$, and $t \bar t$ reduced decay widths $\bar\Gamma$ in the Type I and Type II models and compare to the 2HDM67.

\begin{table}[t]
\begin{center}
\begin{small}
  \begin{tabular}{| c || c | c | c | }
    \hline
$\bar\Gamma(H_\alpha\to f \bar f)$  & Type I & Type II & 2HDM67 \\ \hline\hline
$\bar\Gamma(H_{2,3}\to \tau^+\tau^-)$  &  $\left(\frac{c_\beta}{s_\beta}\right)^2$ & $\left(\frac{s_\beta}{c_\beta}\right)^2$  & $ |\hat\rho^l|^2$ \\ \hline
$\bar\Gamma(H_{2,3}\to b \bar b)$      &  $\left(\frac{c_\beta}{s_\beta}\right)^2$ & $\left(\frac{s_\beta}{c_\beta}\right)^2$  & $ |\hat\rho^d|^2$ \\ \hline
$\bar\Gamma(H_2\to c \bar c)$          &  $\left(\frac{c_\beta}{s_\beta}\right)^2$ & $\left(\frac{c_\beta}{s_\beta}\right)^2$  & $0$ \\ \hline
$\bar\Gamma(H_2\to t \bar t)$          &  $\left(\frac{c_\beta}{s_\beta}\right)^2 (c_3^2\beta_{t\, 2}^2+s_3^2)$ & $\left(\frac{c_\beta}{s_\beta}\right)^2(c_3^2\beta_{t\, 2}^2+s_3^2)$  & $(\Re \hat\rho^u)^2\beta_{t\,2}^2+(\Im\hat\rho^u)^2$ \\ \hline
$\bar\Gamma(H_3\to t \bar t)$          &  $\left(\frac{c_\beta}{s_\beta}\right)^2 (s_3^2\beta_{t\, 2}^2+c_3^2)$ & $\left(\frac{c_\beta}{s_\beta}\right)^2(s_3^2\beta_{t\, 3}^2+c_3^2) $  & $(\Im \hat\rho^u)^2\beta_{t\,3}^2+(\Re\hat\rho^u)^2$ \\ 
    \hline
  \end{tabular}
	\caption{Rescaled decay widths $\bar\Gamma$  for $\tau^+\tau^-$, $b \bar b$, $c \bar c$, and $t \bar t$ final states in the 
	AL. For $\tau$, $b$, and $c$, $\beta_{f\,\alpha}$ was approximated by 1. For the 2HDM67 the assumption of eq.~(\ref{simple_yuk}) was adopted.
	\label{tab:reducedwidth}}
\end{small}
\end{center}
\end{table}

The correlations  between $H_2$ and $H_3$ decay widths are clearly seen from Table~\ref{tab:reducedwidth}. Note first of all that for the light fermions 
the reduced widths are equal for $H_2$ and $H_3$, $\bar\Gamma(H_2\to f \bar f) = \bar\Gamma(H_3\to f \bar f)$, this is a consequence of the alignment and
could be explored to test this scenario. Note however that since it holds also in the Type I and Type II models, it can not be used to disentangle various 
versions of 2HDM5.  The extra freedom provided within the 2HDM67 is also seen from the table, e.g. 
in the Type I and Type II models $\bar\Gamma(H_{2,3}\to \tau^+\tau^-)\simeq\bar\Gamma(H_{2,3} \to b \bar b)$, while in 2HDM67 the reduced widths might be different. 
Note that for the Type I and Type II models leptonic-, down- and up-type widths are correlated while within 2HDM67 they are independent.
Of course, the most straightforward way to disentangle Type I, II and 2HDM67 in the AL is to look for FCNC in $H_{2,3}$ decays, since they are 
not present in the former models while they may appear in the 2HDM67 at the tree level. Of course, the discovery of FCNC by itself would not test the alignment scenario 
as one would need to verify the correlations between the $H_2$ and $H_3$ couplings (and therefore their widths) 
encoded in eqs.~(\ref{sum_rule1})--(\ref{sum_rule2}).

For $t \bar t$ final states the AL might be tested just by measuring $\bar\Gamma_{t\,2,3}$, if a solution with respect to $\Im\hat\rho^u$ and $\Re\hat\rho^u$
exists then the measurement agrees with the AL. Note that the same
measurement could also be interpreted within Type I or Type II.
Then, if eq.~(\ref{magic}) is adopted in Table~\ref{tab:reducedwidth} one obtains for Type I and II the following, $\beta$-dependent relation 
\beq
\frac{\bar\Gamma(H_2\to\bar t t)}{\bar\Gamma(H_3\to\bar t t)}=
\left\{ 
\begin{tabular}{ll}
$\beta_{t\, 2}^2$ & $\alpha_3=0$\\
$\beta_{t\, 3}^{-2}$ & $\alpha_3=\pm \pi/2$
\end{tabular}
\right.
\label{tt_ratio}
\eeq
If $\Gamma_{2,3}$ and masses were measured, then eq.~(\ref{tt_ratio}) could be verified in order to test the alignment scenario for the Type I and II models.

For completeness and future reference we exhibit below the most general Yukawa couplings, expanded up to linear order in $e_{2,3}/v$:
\bea
\bar{l}_kl_lH_1: & &
-\frac{m_k}{v}\delta_{lk}+\frac{1}{2\sqrt{2}v}
\left[
\left(\bar{\rho}^l_{lk}\right)^*(1+\gamma_5)(e_2+ie_3)
+\bar{\rho}^l_{kl}(1-\gamma_5)(e_2-ie_3)
\right],\\
\bar{d}_kd_lH_1:& &
-\frac{m_{d_l}}{v}\delta_{lk}+\frac{1}{2\sqrt{2}v}
\left[\left(\bar{\rho}^d_{lk}\right)^*(1+\gamma_5)(e_2+ie_3)
+\bar{\rho}^d_{kl}(1-\gamma_5)(e_2-ie_3)
\right],\\
\bar{u}_ku_lH_1:& &
-\frac{m_{u_k}}{v}\delta_{lk}+\frac{1}{2\sqrt{2}v}
\left[
\left(\bar{\rho}^u_{lk}\right)^*(1-\gamma_5)(e_2+ie_3)
+\bar{\rho}^u_{kl}(1+\gamma_5)(e_2-ie_3)
\right].
\eea
Below, $\alpha=2$ or $3$:
\bea
\bar{l}_kl_lH_\alpha: & &
- \frac{m_k}{v^2}e_\alpha\delta_{lk}
+\frac{i^\alpha}{2\sqrt{2}}
\left[
\left(\bar{\rho}^l_{lk}\right)^*(1+\gamma_5)
+(-1)^\alpha\bar{\rho}^l_{kl}(1-\gamma_5)
\right],\\
\bar{d}_kd_lH_\alpha: & &
-\frac{m_{d_l}}{v^2}e_\alpha\delta_{lk}
+\frac{i^\alpha}{2\sqrt{2}}
\left[
\left(\bar{\rho}^d_{lk}\right)^*(1+\gamma_5)
+(-1)^\alpha\bar{\rho}^d_{kl}(1-\gamma_5)
\right],\\
\bar{u}_ku_lH_\alpha:& &
-\frac{m_{u_k}}{v^2}e_\alpha\delta_{lk}
+\frac{i^\alpha}{2\sqrt{2}}
\left[
\left(\bar{\rho}^u_{lk}\right)^*(1-\gamma_5)
+(-1)^\alpha\bar{\rho}^u_{kl}(1+\gamma_5)
\right].
\eea
These results show that in the case of a broken $\z2$ symmetry, both flavor-nondiagonal Yukawa couplings of $H_{2,3}$ and new CPV Yukawa couplings are present even in the AL.
Of course, there exist experimental constraints on FCNCs, e.g. measured upper limits for $BR(B_s\to \mu^+\mu^-)$ or 
$B_{s,d}^0$--$\overline{B}\lsup{0}_{s,d}$ mixing that constrain the flavor-nondiagonal Yukawa couplings of the neutral Higgs bosons.  For example, the experimental measurements of the latter
roughly imply that
\beq
\frac{m_b m_k}{M_\alpha^2} \times \left\{ \left|\bar{\rho}^d_{bk}\right|^2, \left|\bar{\rho}^d_{kb}\right|^2, 
\left|\bar{\rho}^d_{bk}\bar{\rho}^{d\, \star}_{kb}\right| \right\} \ll 1 \,,
\eeq
for $k=d,s$ and $\alpha=2,3$. The above constraints would be naturally satisfied in the AL with decoupling for $M_{2,3} \gsim 10\tev$.  In contrast, in the AL  without decoupling with $M_{2,3} \sim {\cal O}(100)\gev$,  the $\bar{\rho}^d_{bk}$ must be sufficiently suppressed.
On the other hand, the presence of some Higgs-mediated FCNCs could be seen as an advantage of the model given that not all FCNC couplings are 
significantly constrained by experiment (as in the case of FCNCs involving the top quark, which could show up in future experimental studies).

%%%%%%%%%%%%%%%%%%%%%%%%%%%%%%%%%%%%%%%%%%%%%%%%%%%%%%%%%%%%%
\subsection{Bosonic decays}
\label{Sec:Bosonic_decays}
%%%%%%%%%%%%%%%%%%%%%%%%%%%%%%%%%%%%%%%%%%%%%%%%%%%%%%%%%%%%%

As we have already seen, fermionic decays of Higgs bosons in the generic model suffer from the presence of
many unknown parameters encoded into the $\rho$ matrices. Therefore it is reasonable to consider only the exact 
AL while investigating fermionic decays. However for bosonic decays we are going to expand the potential
around the alignment up to linear terms in $e_2/v$ and $e_3/v$.
For vector-scalar decay modes we need to expand also the $f_i$ coefficients up to linear order in $e_2/v$ and $e_3/v$
\bea
f_1&=&-\tilde{f}\left(\frac{e_2}{v}-i\frac{e_3}{v}\right),\\
f_2&=&\tilde{f}, \lsp f_3=-i\tilde{f}
\eea
for $\tilde{f}=v(c_3-is_3)$.

\begin{table}[t!]
\begin{center}
\begin{small}
  \begin{tabular}{| c || c | c | c | }
    \hline
operator & exact AL & ${\cal{O}}(e_2/v)$ & ${\cal{O}}(e_3/v)$ \\ \hline\hline
$H_1H_1H_1$ & $M_1^2/(2v)$ &0& 0 \\ \hline
$H_2H_2H_2$ & $\half q_2$ & $(M_2^2-M_{H^\pm}^2)/v$ & $0$ \\ \hline
$H_3H_3H_3$ & $\half q_3$ & $0$ & $(M_3^2-M_{H^\pm}^2)/v$ \\ \hline
$H_1H_1H_2$ & 0 & $(4M_{H^\pm}^2-M_2^2-2vq_1)/(2v)$ & $0$ \\ \hline
$H_1H_1H_3$ &  0 & $0$ & $(4M_{H^\pm}^2-M_3^2-2vq_1)/(2v)$ \\ \hline
$H_2H_2H_1$ & $(2M_2^2-2M_{H^\pm}^2+vq_1)/(2v)$ & $-q_2$& $0$ \\ \hline
$H_3H_3H_1$ & $(2M_3^2-2M_{H^\pm}^2+vq_1)/(2v)$ & $0$ & $-q_3$ \\ \hline
$H_2H_2H_3$ & $\half q_3$ & $0$ & $(M_2^2-M_{H^\pm}^2)/v$ \\ \hline
$H_3H_3H_2$ & $\half q_2$ & $(M_3^2-M_{H^\pm}^2)/v$ & $0$ \\ \hline
$H_1H_2H_3$ & $0$ & $-q_3$ & $-q_2$\\ \hline
$H_iH^-H^+$ & $q_i$ & $0$ & $0$ \\ \hline
 \end{tabular}
	\caption{Coefficients of cubic (non-Goldstone) scalar operators expanded around the alignment limit (AL) \textit{without} decoupling up to 
	${\cal{O}}(e_{2,3}/v)$. The second, third and fourth columns show the exact alignment result and coefficients of 
  $e_2/v$ and $e_3/v$, respectively.  If the alignment limit is realized via decoupling, then certain results of this table are modified as shown in Table~\ref{cubicDL}. \label{cubic}}
\end{small}
\end{center}
\end{table}
\begin{table}[h!]
\begin{center}
\begin{small}
  \begin{tabular}{| c || c | c | c | }
    \hline
operator & exact AL & ${\cal{O}}(e_2/v)$ & ${\cal{O}}(e_3/v)$ \\ \hline\hline
$H_1H_1H_1$ & $M_1^2/(2v)$ & $-e_2 M_{H^\pm}^2/v^2$ & $-e_3M_{H^\pm}^2/v^2$ \\ \hline
$H_1H_1H_2$ & $3e_2M^2_2/(2v^2)$ & $(2M_{H^\pm}^2-2M_2^2-vq_1)/v$ & $0$ \\ \hline
$H_1H_1H_3$ & $3e_3M^2_3/(2v^2)$ & $0$ & $(2M_{H^\pm}^2-2M_3^2-vq_1)/v$ \\ \hline
$H_2H_2H_1$ & $(2M_2^2-2M_{H^\pm}^2+vq_1)/(2v)$ & $-q_2+2e_2 M_2^2/v^2$ & $0$ \\ \hline
$H_3H_3H_1$ & $(2M_3^2-2M_{H^\pm}^2+vq_1)/(2v)$ & $0$ & $-q_3+2e_3 M_3^2/v^2$ \\ \hline
$H_1H_2H_3$ & $0$ & $-q_3+2e_3M_3^2/v^2$ & $-q_2+2e_2M_2^2/v^2$ \\ \hline
 \end{tabular}
	\caption{Coefficients of cubic scalar operators expanded around the alignment limit
	(AL), where the alignment is realized via decoupling, up to 
	${\cal{O}}(e_{2,3}/v)$.   See caption to Table~\ref{cubic}.   	Note that $e_2 M^2$ and
  $e_3 M^2$ (for $M=M_2$, $M_3$ or $M_{H^\pm}$) approach a finite nonzero value in the
  limit of exact decoupling (i.e., as $M\to\infty$). 	
Further explanations are provided in the text. \label{cubicDL}}
\end{small}
\end{center}
\end{table}

Since we are going to focus on two-body Higgs boson decays, in Tables \ref{cubic}--\ref{HHW} we collect coefficients of cubic bosonic operators expanded around the AL up to linear terms in $e_2/v$ and $e_3/v$.  
One subtlety in obtaining the results of Table~\ref{cubic} is the distinction between achieving the alignment limit via decoupling or in the absence of decoupling.  We illustrate this point by examining the $H_1 H_1 H_2$ coupling.  \Eq{hhh2} yields the following coefficient of the $H_1 H_1 H_2$ operator in the scalar potential,
\beq \label{h1h1h2coeff}
H_1 H_1 H_2:\quad -\frac{e_1 e_2}{v^2}q_1+\frac{v^2-e_1^2}{2v^2}q_2+\frac{(3e_1^2-v^2)e_2}{v^4}M^2_{H^\pm}+\frac{(v^2-e_1^2)e_2}{v^4}M_1^2-\frac{e_1^2 e_2}{2v^4}M_2^2\,.
\eeq
In the approximate alignment limit \textit{without} decoupling, all scalar squared masses are of $\mathcal{O}(v^2)$.    In light of Table~\ref{tabal},
the coefficient of the $H_1 H_1 H_2$ operator is
\beq \label{h112op}
v\left[\frac{e_2(4M_{H^\pm}^2-M_2^2-2vq_1)}{2v^3}+\mathcal{O}(e_2^2/v^2,e_3^2/v^2)\right]\,,
\eeq
where we have explicitly exhibited the terms of $\mathcal{O}(e_2/v)$ inside the bracketed expression above [note that there are no terms of $\mathcal{O}(e_3/v)$].  In the exact alignment limit (where we set $e_2=e_3=0$), the coefficient of the $H_1 H_1 H_2$ operator vanishes.
In contrast, in the decoupling regime, $M^2_2$, $M_{H^\pm}^2\gg v^2$, and the expansion in the small parameters 
is organized differently.   In particular, using the results of Appendix~\ref{app:al}, one can derive \eq{app:decouplim}, which yields,
\beq \label{decouplim}
e_2 M^2 \simeq v^3\Re(Z_6 e^{-i\theta_{23}})\,,\qquad\quad e_3 M^2 \simeq -v^3\Im(Z_6 e^{-i\theta_{23}})\,,
\quad \text{for $M=M_2, M_3, M_{H^\pm}$}\,,
\eeq
where $Z_6$ is an $\mathcal{O}(1)$ parameter that appears in the scalar potential expressed in terms of the Higgs basis fields [cf.~\eq{higgspot}], and $\theta_{23}$ is a mixing angle introduced in \eq{mixingmatrix}.
Hence, in the exact alignment limit in the decoupling regime, \eq{h112op} yields $3e_2 M^2_2/(2v^2)$, which is finite and nonzero as $M_2\to\infty$ in light of \eq{decouplim}.  The first order correction to this result is
\beq 
\frac{e_2\bigl[2(M_{H^\pm}^2-M_2^2)-vq_1\bigr]}{v^2}\,,
\eeq
where the difference in the squared masses above is given by \eq{massdiff2}.  Similar considerations apply to the $H_1 H_1 H_3$ operator in 
Table~\ref{cubic}.

Similarly, in the $\mathcal{O}(e_2/v)$ and $\mathcal{O}(e_3/v)$ entries for the $H_1 H_1 H_1$, $H_2 H_2 H_1$, $H_3 H_3 H_1$ and $H_1 H_2 H_3$ operators, the results presented in Table~\ref{cubic} do not include terms that
are of $\mathcal{O}(e^2_2 M^2/v^3)$ and $\mathcal{O}(e^3_2 M^2/v^3)$, where $M=M_2, M_3$ or  $M_{H^\pm}$.
Such terms are quadratically suppressed in the approximate alignment limit without decoupling.  But in the decoupling regime, \eq{decouplim} implies that such terms would compete with those terms listed in Table~\ref{cubic}.   Thus, in Table~\ref{cubicDL}, we provide the exact alignment results and the corresponding first order corrections for those cubic Higgs operators that differ from the results displayed in Table~\ref{cubic}.

\begin{table}[t!]
\begin{center}
\begin{small}
  \begin{tabular}{| c || c | c | c | }
    \hline
operator & exact AL & ${\cal{O}}(e_2/v)$ & ${\cal{O}}(e_3/v)$ \\ \hline\hline
$H_1 H^+ W^-_\mu$ & $0$ & $-\frac{ig}{2v} \tilde{f} (p_1-p^+)_\mu$       & $-\frac{g}{2v} \tilde{f} (p_1-p^+)_\mu$ \\ \hline
$H_1 H^- W^+_\mu$ & $0$ & $+\frac{ig}{2v} \tilde{f}^\ast (p_1-p^-)_\mu$  & $-\frac{g}{2v} \tilde{f}^\ast (p_1-p^-)_\mu$ \\ \hline
$H_2 H^+ W^-_\mu$ & $+\frac{ig}{2v} \tilde{f} (p_2-p^+)_\mu$ & $0$       & $0$ \\ \hline
$H_2 H^- W^+_\mu$ & $-\frac{ig}{2v} \tilde{f}^\ast (p_2-p^-)_\mu$   & $0$       & $0$ \\ \hline
$H_3 H^+ W^-_\mu$ & $+\frac{g}{2v} \tilde{f} (p_3-p^+)_\mu$ & $0$       & $0$ \\ \hline
$H_3 H^- W^+_\mu$ & $+\frac{g}{2v} \tilde{f}^\ast (p_3-p^-)_\mu$   & $0$       & $0$ \\ \hline
$H_1 H_2 Z_\mu$ & $0$ & $0$ & $\frac{g}{2\cos\theta_W} (p_1-p_2)_\mu$ \\ \hline
$H_2 H_3 Z_\mu$ & $\frac{g}{2\cos\theta_W} (p_2-p_3)_\mu$ & $0$ & $0$ \\ \hline
$H_3 H_1 Z_\mu$ & $0$ & $\frac{g}{2\cos\theta_W} (p_3-p_1)_\mu$ & $0$ \\ \hline
$H_1 Z_\mu Z_\nu$ & $\frac{ig^2v}{2\cos^2\theta_W}g_{\mu\nu}$ & $0$ & $0$ \\ \hline
$H_2 Z_\mu Z_\nu$ & $0$ & $\frac{ig^2v}{2\cos^2\theta_W}g_{\mu\nu}$ & $0$ \\ \hline
$H_3 Z_\mu Z_\nu$ & $0$ & $0$ & $\frac{ig^2v}{2\cos^2\theta_W}g_{\mu\nu}$ \\ \hline
$H_1 W^+_\mu W^-_\nu$ & $\frac{ig^2v}{2}g_{\mu\nu}$ & $0$ & $0$ \\ \hline
$H_2 W^+_\mu W^-_\nu$ & $0$ & $\frac{ig^2v}{2}g_{\mu\nu}$ & $0$ \\ \hline
$H_3 W^+_\mu W^-_\nu$ & $0$ & $0$ & $\frac{ig^2v}{2}g_{\mu\nu}$ \\ \hline
 \end{tabular}
	\caption{Coefficients of $H_i H^+ W^-_\mu$, $H_i H^- W^+_\mu$, $H_i H_j Z_\mu$, $H_iZ_\mu Z_\nu$ and $H_iW^+_\mu W^-_\nu$  operators expanded around the AL up to ${\cal{O}}(e_{2,3}/v)$. The second, third and fourth columns show the alignment result and coefficients of 
  $e_2/v$ and $e_3/v$, respectively. \label{HHW}}
\end{small}
\end{center}
\end{table}
\begin{table}[h!]
\begin{center}
\begin{small}
  \begin{tabular}{| c || c | }
    \hline
$\Gamma$ & decay process \\ \hline\hline
${\cal{O}}(1)$ & $H^\pm\to H_{2,3} W^\pm$, $H_{2,3}\to H^+H^-$, $H_{2,3}\to H_1 H_1$ (DL), $H_3\to H_2H_2$, $H_3\to H_2 Z$ \\ \hline
${\cal{O}}\big[\left(\frac{e_2}{v}\right)^2\big]$ & $H_3\to H_1 Z$, $H_2\to ZZ$, $H_2\to W^+W^-$, $H_2\to H_1 H_1$\\ \hline
${\cal{O}}\big[\left(\frac{e_3}{v}\right)^2\big]$ & $H_2\to H_1 Z$, $H_3\to ZZ$, $H_3\to W^+W^-$, $H_3\to H_1 H_1$\\ \hline
${\cal{O}}\big[\left(\frac{e_2}{v}\right)^2,\left(\frac{e_3}{v}\right)^2,\frac{e_2e_3}{v^2}\big]$ & $H_3\to H_2 H_1$\\ \hline
\end{tabular}
	\caption{Possible two-body decays of heavy Higgs bosons classified according to the strength of the corresponding decay width. 
	The first row shows leading decays that exist in the AL, the second and third rows show decays the width of which is suppressed by $(e_2/v)^2$ and $(e_3/v)^2$, respectively, while the fourth one contains decays with the width suppressed by $\max[(e_2/v)^2, (e_3/v)^2, (e_2e_3/v^2)]$. As noted in the text [cf.~Tables~\ref{cubic} and \ref{cubicDL}], if approximate alignment is achieved in the decoupling limit (DL), then the coefficient of the $H_1 H_1 H_k$ operator (for $k=2,3$) in the exact alignment limit is nonzero and hence unsuppressed.\label{decays}
}
\end{small}
\end{center}
\end{table}

In Table~\ref{decays} we combine predictions for bosonic Higgs boson two-body decays.
Note that ratios of decay widths for processes contained in the second and third rows of Table~\ref{decays} are functions of masses of the involved particles only, as the couplings proportional to $e_2$ and $e_3$ cancel out. Therefore, even though the widths are expected to be small, predictions for their ratios are quite unambiguous, depending on masses of the involved particles only.   

%%%%%%%%%%%%%%%%%%%%%%%%%%%%%%%%%%%%%%%%%%%%%%%%%%%%%%%%%%%%%
\subsubsection{Scalar-scalar modes}

Among the leading, unsuppressed decays contained in the first row of Table~\ref{decays} only two purely scalar decays $H_3\to H^+H^-$, $H_3\to H_2H_2$ are subject of some extra uncertainties as they are both $\propto q_3$, however their ratio
\beq
\frac{\BR(H_3\to H^+H^-)}{\BR(H_3\to H_2H_2)} 
= \sqrt{\frac{M_3^2-4M_{H^\pm}^2}{M_3^2-4M_2^2}}
\left[1-\frac{4(M_2^2-M_{H^\pm}^2)}{q_3v}\frac{e_3}{v}
+\mathcal{O}(e_2^2/v^2,e_2e_3/v^2,e_3^2/v^2)\right],
\label{rel1}
\eeq
depends only on the masses of the particles involved in the decays and therefore might be useful for testing this scenario.
Note that in the 2HDM5 models $H_3\to H^+H^-$ and $H_3\to H_2H_2$ decays exist only if $\alpha_3=\pm\pi/2$ 
($A=H_2$ and $H=H_3$). In the case
of $\alpha_3=0$ ($A=H_3$ and $H=H_2$) the corresponding widths vanish at the tree level.

Furthermore, as seen from Table~\ref{decays}, we find that to the leading order
\beq
\BR(H_3 \to H_1 H_2) = 
	\mathcal{O}(e_2^2/v^2,e_2e_3/v^2,e_3^2/v^2).
\eeq
In contrast, the decay rate for $H_{2,3}\to H_1 H_1$ may or may not be suppressed depending on whether approximate alignment is achieved with or without decoupling.   Indeed as previously noted, the leading contribution to the $H_{k}H_1H_1$ coupling (for $k=2,3$) is proportional to  $e_k M_k^2/v^2\sim \mathcal{O}(v)$ in the limit of large $M_k\gg v$ and hence unsuppressed.

%%%%%%%%%%%%%%%%%%%%%%%%%%%%%%%%%%%%%%%%%%%%%%%%%%%%%%%%%%%%%
\subsubsection{Scalar-vector modes}

In this subsection, we consider Higgs decay modes into two-body $VV$ and $VH$ final states (for $V=W^\pm$ or $Z$ and $H=H_{1,2,3},H^\pm$).\footnote{In the CP-conserving 2HDM, the phenomenology of these decay modes are discussed in Refs.~\cite{Coleppa:2014hxa,Coleppa:2014cca,Li:2015lra,Kling:2016opi}.}
In the AL, $H_1$ couples to the vector bosons as in the SM, whereas
the decay rates for the modes $H_{2,3}\to W^+W^-$, $H_{2,3}\to ZZ$ and $H_{2,3}\to H_1 Z$ all vanish, since their couplings are
proportional to $e_{2,3}=0$.  Hence, it follows that
\bea
\frac{\Gamma(H_1 \to W^+W^-,ZZ)}{\Gamma(H_\text{SM} \to W^+W^-,ZZ)} &=&1
	+\mathcal{O}(e_2^2/v^2,e_2e_3/v^2,e_3^2/v^2), \\ 
\BR(H_{2,3} \to W^+W^-, ZZ, H_1 Z) &=& 
	\mathcal{O}(e_2^2/v^2,e_2e_3/v^2,e_3^2/v^2).
\eea
The couplings $H_2 H^+W^-$ and  $H_3 H^+W^-$ differ by a phase factor only,
so that (if kinematically open)\footnote{The K\"{a}ll\'{e}n function is defined as $\lambda(x,y,z)=x^2+y^2+z^2-2xy-2xz-2yz$.}
\beq
\frac{\BR(H^\pm \to H_2W^\pm)}{\BR(H^\pm \to H_3W^\pm)} 
= \frac{\left[\lambda(M_{H^\pm},M_2,M_W)\right]^{3/2}}{\left[\lambda(M_{H^\pm},M_3,M_W)\right]^{3/2}}
+\mathcal{O}(e_2^2/v^2,e_2e_3/v^2,e_3^2/v^2).
\label{rel2} 
\eeq
Similarly, the couplings $H_3 H^+W^-$ and $H_3H_2Z$ differ by a phase and a trivial factor only, so
\beq
\frac{\BR(H_3 \to H_2Z)}{\BR(H_3 \to H^+W^-)} = \frac{1}{c_W^2}
\frac{\left[\lambda(M_3,M_2,M_Z)\right]^{3/2}}{\left[\lambda(M_3,M_{H^\pm},M_W)\right]^{3/2}}
+	\mathcal{O}(e_2^2/v^2,e_2e_3/v^2,e_3^2/v^2). 
\label{rel3}
\eeq
The observables defined above do not differentiate between the 2HDM5 and 2HDM67 models. 

Since in the AL $f_1=0$ therefore 
\beq
\BR(H^\pm \to W^\pm H_1) = 
	\mathcal{O}(e_2^2/v^2,e_2e_3/v^2,e_3^2/v^2).
\label{rel4}
\eeq  
This holds both for the 2HDM5 and 2HDM67 models.

Note that for the $VV$ and $VH$ final states discussed in this subsection, the leading corrections to the AL results appear at the quadratic level, i.e., there are no corrections linear in $e_{2,3}$. Therefore the above predictions in the AL are rather robust.

%%%%%%%%%%%%%%%%%%%%%%%%%%%%%%%%%%%%%%%%%%%%%%%%%%%%%%%%%%%%%
\section{Summary}
\label{Sec:summary}
\setcounter{equation}{0}
%%%%%%%%%%%%%%%%%%%%%%%%%%%%%%%%%%%%%%%%%%%%%%%%%%%%%%%%%%%%

Given that the observed couplings of the Higgs boson are SM-like and the $\rho$ parameter is measured to be near $1$,
models of extended Higgs sectors are significantly constrained.  The generic two Higgs doublet model (2HDM) unconstrained by a $\z2$ symmetry
provides a simple extension of the SM with new sources of CP violation in the scalar sector.
In this paper, 
the phenomenology of the approximate alignment limit of the 2HDM in which the $125\gev$ Higgs boson couplings are close to those of the SM 
has been discussed in detail. 
The alignment limit can be achieved with or without the decoupling of the heavier Higgs states of the scalar sector.  Indeed, 
regions of the 2HDM parameter space exist in which at least some of the heavier scalar states have masses not significantly above the observed Higgs boson mass of 
$125\gev$, 
without being in conflict with the SM-like couplings of the discovered Higgs boson.\footnote{A comprehensive analysis of the generic 2HDM parameter space consistent with all existing experimental observables will be postponed for future work.  In this paper, we simply emphasize that even if the additional scalars of the 2HDM are light,
the structure of the model does not preclude the SM-like Higgs boson coupling from being SM-like.} 

We have shown that all possible bosonic couplings of the 2HDM scalars can be expressed in terms of a minimal set of seven physical Higgs couplings and four scalar masses,
which then yield numerous correlations among processes involving the interactions of the scalars.
Some of these correlations are quite striking in the alignment limit;  for example, 
$\BR(H_{2,3} \to W^+W^-, ZZ, ZH_1) =0$ and  
$\BR(H^\pm \to W^\pm H_1) = 0$.  In addition, correlations between $H_2$ and $H_3$ couplings in the alignment limit imply that
the ratios of branching ratios, $\BR(H^\pm \to H_2W^\pm)/\BR(H^\pm \to H_3W^\pm)$, $\BR(H_3 \to H_2Z)/\BR(H_3 \to H^+W^-)$ for bosonic decays and
$\Gamma(H_2 \to f\bar f)/\Gamma(H_3 \to f\bar f)$ for fermionic ones, are functions of masses only.
Leading corrections ($\propto e_{2,3}$) to the alignment limit results for bosonic decays of $H_{2,3}$ have been also calculated.  In particular, for the scalar-vector final states, the corrections to exact alignment are quadratic in small quantities, i.e. they are proportional to $e_2^2, e_2e_3$ or $e_3^2$.  Consequently, the alignment limit results for branching ratios into scalar-vector final states are quite robust.

In processes that involve the cubic or quartic scalar couplings, the implications of the alignment limit may depend on whether alignment is achieved via the decoupling of heavy scalar states.   For example, we have shown that in the exact alignment limit without decoupling, BR($H_{2,3}\to H_1 H_1)=0$.  In contrast, in the decoupling regime, this branching ratio is finite and non-zero. 

For the Yukawa couplings, in addition to presenting the most general results, various special cases were considered, e.g. diagonal $\rho$ matrices, type I or II models, etc. Leading corrections to the exact alignment limit values have been also shown and expressed in terms of the (small, i.e. $\propto e_{2,3}$) couplings of the heavier scalars to the gauge bosons. It should be stressed that broken $\z2$ implies not only new sources of CP-violation in the potential, but also extra CP-violating Yukawa couplings. Phenomenologically the latter might be even more relevant in future collider studies.

The absence of a $\z2$ symmetry, even in the alignment limit, results in the presence of flavor-nondiagonal Yukawa couplings of the heavier neutral Higgs bosons. For some of them there exist severe experimental 
upper limits which would be satisfied in the alignment limit for sufficiently heavy non-SM-like Higgs boson masses (i.e., in the alignment limit with decoupling).  In contrast, the alignment limit without decoupling with $M_{2,3}\sim {\cal O}(100)\gev$ requires a significant fine-tuning of the flavor-nondiagonal couplings to be consistent with experimental constraints.  In this context, we note that current experimental constraints on flavor-nondiagonal neutral Higgs coupling to top quarks are quite weak.  If evidence of such couplings emerge in future experiments, such phenomena could be easily accommodated in the generic 2HDM.

%%%%%%%%%%%%%%%%%%%%%%%%%%%%%%%%%%%%%%%%%%%%%%%%%%%%%%%%%%%%
\section*{Acknowledgments}
B.G. acknowledges partial support by the National Science Centre (Poland) research project no 2017/25/B/ST2/00191. 
H.E.H. is supported in part by the U.S. Department of Energy grant
number DE-SC0010107, and in part by the grant H2020-MSCA-RISE-2014
No. 645722 (NonMinimalHiggs).
This work is also supported in part by the National Science Centre (Poland) HARMONIA project under contract UMO-2015/18/M/ST2/00518 (2016-2019).
P.O. is supported in part by the Research Council of Norway.

\appendix
%%%%%%%%%%%%%%%%%%%%%%%%%%%%%%%%%%%%%%%%%%%%%%%%%%%%%%%%%%%%
\section{Couplings involving gauge fields}
\label{sect:cubic_gauge_couplings}
%%%%%%%%%%%%%%%%%%%%%%%%%%%%%%%%%%%%%%%%%%%%%%%%%%%%%%%%%%%%
By putting
\bea
D^\mu&=&\partial^\mu+\frac{ig}{2}\sigma_iW_i^\mu+i\frac{g^\prime}{2}B^\mu,
\eea
with $W_1^\mu=\frac{1}{\sqrt{2}}(W^{+\mu}+W^{-\mu})$, $W_2^\mu=\frac{i}{\sqrt{2}}(W^{+\mu}-W^{-\mu})$, $W_3^\mu=\cos\thetaW Z^\mu+\sin\thetaW A^\mu$ and $B^\mu=-\sin\thetaW Z^\mu+\cos\thetaW A^\mu$,
the kinetic part of the Lagrangian can be written
\begin{equation} \label{Eq:gauge-IS}
\lcal_k=(D_\mu \Phi_1)^\dagger(D^\mu \Phi_1) + (D_\mu \Phi_2)^\dagger(D^\mu \Phi_2).
\end{equation}
From $\lcal_k$ we can now read off directly coefficients representing the interactions involving both scalars and vector bosons:
%%%%%%%%%%%%%%%%%%%%%%%%%%%%%%%%%%%%%%%%%%%%%%%%%%%%%%%%%%%%
\subsection{Trilinear couplings involving one scalar and two vector bosons}
%%%%%%%%%%%%%%%%%%%%%%%%%%%%%%%%%%%%%%%%%%%%%%%%%%%%%%%%%%%%
By reading off the coefficients from the kinetic part of the Lagrangian we find\footnote{In order to promote these coefficients to Feynman rules we should multiply with $i$ and an appropriate combinatorial factor if the vertex contains identical particles.}
\begin{subequations}
	\begin{alignat}{2}
	H_i Z_\mu Z^\mu:\quad &
	\frac{g^2}{4\cos^2\thetaW}e_i, &\quad
	H_i W^+_\mu W^{-\mu}:\quad &
	\frac{g^2}{2}e_i,  \\
	G^\pm W^{\mp}_\mu A^\mu: \quad & \frac{g^2v}{2}\sin\theta_W, &\quad
	G^\pm W^{\mp}_\mu Z^\mu: \quad & -\frac{g^2v}{2}\frac{\sin^2\theta_W}{\cos\theta_W}.
	\end{alignat}
\end{subequations}
The factors $e_i$ parametrizing the first two of these couplings play an important role. They are given by
\bea
e_i\equiv v_1R_{i1}+v_2R_{i2},
\eea
and are known to be basis invariant quantities.
%%%%%%%%%%%%%%%%%%%%%%%%%%%%%%%%%%%%%%%%%%%%%%%%%%%%%%%%%%%%
\subsection{Trilinear couplings involving two scalars and one vector boson}
%%%%%%%%%%%%%%%%%%%%%%%%%%%%%%%%%%%%%%%%%%%%%%%%%%%%%%%%%%%%
The coefficients of the kinetic part of the Lagrangian are in these cases found to be\footnote{Eq. (\ref{eq:misprint}) corrects a misprint in eq.~(B.25d) of Ref.~\cite{Grzadkowski:2014ada}.}
\begin{subequations}
	\begin{alignat}{2}
(H_i\ddel_\mu\!H_j )Z^\mu: \quad &
-\frac{g}{2v\cos\thetaW}\epsilon_{ijk}e_k, &\quad
(G^0\ddel_\mu\!H_i )Z^\mu: \quad &
\frac{g}{2v\cos\thetaW}e_i,\\
(H^+\ddel_\mu\!H^- )A^\mu: \quad &
ig\sin\theta_W, &\quad
(H^+\ddel_\mu\!H^- )Z^\mu: \quad &
i\frac{g}{2}\frac{\cos2\theta_W}{\cos\theta_W},\\
(G^+\ddel_\mu\!G^- )A^\mu: \quad &
ig\sin\theta_W, &\quad
(G^+\ddel_\mu\!G^- )Z^\mu: \quad &
i\frac{g}{2}\frac{\cos2\theta_W}{\cos\theta_W},\\
(G^\pm\ddel_\mu\!H_i )W^{\mp\mu}: \quad &
\pm i\frac{g}{2v}e_i, &\quad
(G^\pm\ddel_\mu\!G^0 )W^{\mp\mu}: \quad &
\frac{g}{2},\\
(H^+\ddel_\mu\!H_i )W^{-\mu}: \quad &
i\frac{g}{2v}f_i, &\quad
(H^-\ddel_\mu\!H_i )W^{+\mu}: \quad &
-i\frac{g}{2v}f^*_i\label{eq:misprint}.
	\end{alignat}
\end{subequations}
Here, we encounter the coefficients $f_i$ (and their conjugate partners $f_i^*$),  which appear in couplings between scalars and gauge bosons whenever an $H^+W^-$ pair ($H^-W^+$ pair) is present in the vertex\footnote{As we shall soon see, $f_i$ and $f_i^*$ also appear in scalar couplings whenever an $H^+G^-$ or $H^-G^+$ pair is present.}.
These factors are defined by
\begin{equation}
f_i\equiv v_1R_{i2}-v_2R_{i1}-ivR_{i3},
\end{equation}
and satisfy the relation given in eq.~(\ref{fifj}).
%%%%%%%%%%%%%%%%%%%%%%%%%%%%%%%%%%%%%%%%%%%%%%%%%%%%%%%%%%%%
\subsection{Quadrilinear couplings involving two scalars and two vector bosons}
%%%%%%%%%%%%%%%%%%%%%%%%%%%%%%%%%%%%%%%%%%%%%%%%%%%%%%%%%%%%
For the coefficients of the quadrilinear couplings we find:
\begin{subequations}
	\begin{alignat}{2}
	H_i H_i Z_\mu Z^\mu: \quad &
	\frac{g^2}{8\cos^2\theta_W}, &\quad
	G^0 G^0 Z_\mu Z^\mu: \quad &
	\frac{g^2}{8\cos^2\theta_W}, \\
	H_i H_i W^+_\mu W^{-\mu}: \quad &
	\frac{g^2}{4}, &\quad
	G^0 G^0 W^+_\mu W^{-\mu}: \quad &
	\frac{g^2}{4}, \\
	H_i G^\pm A_\mu W^{\mp\mu}: \quad &
	\frac{g^2}{2v}\sin\thetaW e_i , &\quad
	G^0 G^\pm A_\mu W^{\mp\mu}: \quad &
	\mp i\frac{g^2}{2}\sin\thetaW, \\
	G^- G^+ A_\mu A^\mu: \quad &
	g^2 \sin^2\theta_W, &\quad
	H^- H^+ A_\mu A^\mu: \quad &
	g^2 \sin^2\theta_W, \\
	G^- G^+ A_\mu Z^\mu: \quad &
	g^2\tan\theta_W \cos2\theta_W, &\quad
	H^- H^+ A_\mu Z^\mu: \quad &
	g^2\tan\theta_W \cos2\theta_W, \\
	G^- G^+ Z_\mu Z^\mu: \quad &
	\frac{g^2}{4}\frac{\cos^22\theta_W}{\cos^2\theta_W}, &\quad
	H^- H^+ Z_\mu Z^\mu: \quad &
	\frac{g^2}{4}\frac{\cos^22\theta_W}{\cos^2\theta_W}, \\
	G^- G^+ W^+_\mu W^{-\mu}: \quad &
	\frac{g^2}{2}, &\quad
	H^- H^+ W^+_\mu W^{-\mu}: \quad &
	\frac{g^2}{2}, \\
	H_i G^\pm Z_\mu W^{\mp\mu}: \quad &
	-\frac{g^2}{2v} \frac{\sin^2\thetaW}{\cos\thetaW}e_i, &\quad
	G^0 G^\pm Z_\mu W^{\mp\mu}: \quad &
	\pm i\frac{g^2}{2} \frac{\sin^2\theta_\text{W}}{\cos\thetaW},\\
	H_i H^+ A_\mu W^{-\mu}: \quad &
	\frac{g^2}{2v} \sin\theta_Wf_i, &\quad
	H_i H^- A_\mu W^{+\mu}: \quad &
	\frac{g^2}{2v} \sin\theta_Wf^*_i,\\
	H_i H^+ Z_\mu W^{-\mu}: \quad &
	-\frac{g^2}{2v} \frac{\sin^2\theta_W}{\cos\theta_W}f_i, &\quad
	H_i H^- Z_\mu W^{+\mu}: \quad &
	-\frac{g^2}{2v} \frac{\sin^2\theta_W}{\cos\theta_W}f^*_i.
	\end{alignat}
\end{subequations}

%%%%%%%%%%%%%%%%%%%%%%%%%%%%%%%%%%%%%%%%%%%%%%%%%%%%%%%%%%%%
\section{Cubic coefficients from the potential}
\label{sect:cubic_couplings}
%%%%%%%%%%%%%%%%%%%%%%%%%%%%%%%%%%%%%%%%%%%%%%%%%%%%%%%%%%%%
The trilinear couplings\footnote{We present here the coefficients of the potential. In order to promote these to Feynman rules one must multiply by $-i$ due to the fact that the potential appears with a negative sign in the Lagrangian as well as an appropriate combinatorial factor if the vertex contains identical particles.} among the scalars can be expressed in terms of the eleven observables (masses/couplings) of $\pcal$, eq.~(\ref{Eq:pcal})  (and the auxiliary quantities $f_i$) as follows:
\begin{alignat}{2}
&H_iH_iH_i:&\quad&
\frac{v^2-e_i^2}{2v^2}q_i-\frac{(v^2-e_i^2)e_i}{v^4}M_{H^\pm}^2
+\frac{(2v^2-e_i^2)e_i}{2v^4}M_i^2 \label{hhh1}\\
&H_iG^0G^0:&\quad&\frac{e_i}{2v^2}M_i^2 \\
&H_iH_iH_j:&\quad&
-\frac{e_ie_j}{v^2}q_i+\frac{v^2-e_i^2}{2v^2}q_j
+\frac{(3e_i^2-v^2)e_j}{v^4}M_{H^\pm}^2
+\frac{(v^2-e_i^2)e_j}{v^4}M_i^2
-\frac{e_i^2e_j}{2v^4}M_j^2 \label{hhh2} \\
&G^0H_jH_j:&\quad &0\\
&G^0H_iH_j:&\quad&\frac{1}{v^2}\sum_{k}\epsilon_{ijk}e_k(M_i^2-M_j^2) \\
&H_1H_2H_3:&\quad&
-\frac{e_2e_3}{v^2}q_1-\frac{e_1e_3}{v^2}q_2-\frac{e_1e_2}{v^2}q_3
+\frac{6e_1e_2e_3}{v^4}M_{H^\pm}^2-\frac{e_1e_2e_3}{v^4}(M_1^2+M_2^2+M_3^2) \label{hhh3} \\
&H_iG^+G^-:&\quad&\frac{e_i}{v^2}M_i^2\\
&H_iH^+H^-:&\quad&q_i \\
&H_iH^+G^-:&\quad&
\frac{f_i}{v^2}(M_i^2-M_{H^\pm}^2)
\end{alignat}
Charge-conjugated vertices are related by complex conjugation, $H_iG^+H^-=(H_iH^+G^-)^\ast$.

%%%%%%%%%%%%%%%%%%%%%%%%%%%%%%%%%%%%%%%%%%%%%%%%%%%%%%%%%%%%
\section{Quartic coefficients from the potential}
\label{sect:quartic_couplings}
%%%%%%%%%%%%%%%%%%%%%%%%%%%%%%%%%%%%%%%%%%%%%%%%%%%%%%%%%%%%
The quartic couplings among the scalars can be expressed in terms of the $\cal{P}$ parameters of eq.~ (\ref{Eq:pcal}) supplemented by $f_i$ as follows:

\begin{alignat}{2}
&G^0G^0G^0G^0:&\quad&\frac{1}{8 v^4}\left(e_1^2M_1^2+e_2^2M_2^2+e_3^2M_3^2\right)\\
&H_iH_iH_iH_i:&\quad
&\frac{(v^2-e_i^2)^2}{4v^4}q
+\frac{(v^2-e_i^2)e_i}{2v^4}q_i
+\frac{e_i^4}{8v^8}(e_1^2M_1^2+e_2^2M_2^2+e_3^2M_3^2)
\nonumber \\
&\quad&&
-\frac{(v^2-e_i^2)e_i^2}{4v^6}(e_1q_1+e_2q_2+e_3q_3 + 2M_{H^\pm}^2 -2M_i^2) \label{hhhh1} \\
&H_iG^0G^0G^0:&\quad&\frac{1}{4 v^4}
\sum_{j,k}\epsilon_{ijk}e_je_k(M_k^2-M_j^2) \\
&H_iH_iH_iH_j:&\quad&
\frac{(e_i^2-v^2)e_ie_j}{v^4}q
+\frac{(v^2-3e_i^2)e_j}{2v^4}q_i
+\frac{(v^2-e_i^2)e_i}{2v^4}q_j \nonumber\\
&\quad&&
+\frac{(2e_i^2-v^2)e_ie_j}{2v^6}(e_1q_1+e_2q_2+e_3q_3 + 2M_{H^\pm}^2 )
\nonumber \\
&\quad&&
+\frac{(2v^2-3e_i^2)e_ie_j}{2v^6}M_i^2
-\frac{e_i^3e_j}{2v^6}M_j^2
+\frac{e_i^3e_j}{2v^8}(e_1^2M_1^2+e_2^2M_2^2+e_3^2M_3^2) \label{hhhh2}\\
&G^0H_iH_iH_i:&\quad&
\frac{v^2-e_i^2}{2v^4}\sum_{j,k}\epsilon_{ijk}e_jq_k
+\frac{e_i^2}{2v^6}\sum_{j,k}\epsilon_{ijk}e_je_kM_j^2 \\
&H_iH_iG^0G^0:&\quad&
\frac{v^2-e_i^2}{4v^4}(e_1q_1+e_2q_2+e_3q_3-2M_{H^\pm}^2
+2M_1^2+2M_2^2+2M_3^2)
+\frac{3e_i^2-v^2}{2v^4}M_i^2 \nonumber \\
&\quad&&
-\frac{2v^2+e_i^2}{4v^6}(e_1^2M_1^2+e_2^2M_2^2+e_3^2M_3^2)\\
&H_iH_iH_jH_j:&\quad&\frac{v^4-(e_i^2+e_j^2)v^2+3e_i^2e_j^2}{2v^4}q
+\frac{v^2-3e_j^2}{2v^4}e_iq_i
+\frac{v^2-3e_i^2}{2v^4}e_jq_j  \nonumber \\
&\quad&&
+\frac{6e_i^2e_j^2-(e_i^2+e_j^2)v^2}{4v^6}(e_1q_1+e_2q_2+e_3q_3 +2M_{H^\pm}^2)
+\frac{(v^2-3e_i^2)e_j^2}{2v^6}M_i^2
\nonumber \\
&\quad&\quad&
+\frac{(v^2-3e_j^2)e_i^2}{2v^6}M_j^2
+\frac{3e_i^2e_j^2}{4v^8}(e_1^2M_1^2+e_2^2M_2^2+e_3^2M_3^2) \label{hhhh3} \\
&H_iH_jG^0G^0:&\quad&-\frac{e_ie_j}{2v^4}(e_1q_1+e_2q_2+e_3q_3-2M_{H^\pm}^2
+2M_1^2+2M_2^2+2M_3^2) \nonumber\\
&\quad&&
+\frac{3e_ie_j}{2v^4}(M_i^2+M_j^2)
-\frac{e_ie_j}{2v^6}(e_1^2M_1^2+e_2^2M_2^2+e_3^2M_3^2) \\
&H_iH_iH_jH_k:&\quad&
\frac{(3e_i^2-v^2)e_je_k}{v^4}q
+\frac{(v^2-3e_i^2+6e_j^2)e_k}{2v^4}q_j
+\frac{(v^2-3e_i^2+6e_k^2)e_j}{2v^4}q_k  \nonumber\\
&\quad&&
+\frac{(6e_i^2-7v^2)e_je_k}{2v^6}(e_1q_1+e_2q_2+e_3q_3)
+\frac{(6e_i^2-v^2)e_je_k}{v^6}M_{H^\pm}^2\nonumber\\
&\quad&\quad&
+\frac{(2v^2-3e_i^2)e_je_k}{2v^6}M_i^2
-\frac{3e_i^2e_je_k}{2v^6}(M_1^2+M_2^2+M_3^2) \nonumber \\
&\quad&\quad&
+\frac{3e_i^2e_je_k}{2v^8}(e_1^2M_1^2+e_2^2M_2^2+e_3^2M_3^2) \label{hhhh4} \\
&G^0H_iH_jH_j:&\quad&
\left[-\frac{e_ie_j}{v^4}q_i-\frac{(v^2-e_j^2)}{2v^4}q_j\right]
\sum_k \epsilon_{ijk}e_k
+\frac{(v^2+2e_i^2-e_j^2)e_j}{2v^4}\sum_k \epsilon_{ijk}q_k \\
&\quad&&
+\left[\frac{(v^2-e_i^2)e_j}{v^6}M_i^2
+\frac{(e_j^2-2v^2)e_j}{2v^6}M_j^2\right]\sum_k \epsilon_{ijk}e_k
+\frac{(2e_i^2-e_j^2)e_j}{2v^6}\sum_k \epsilon_{ijk}e_kM_k^2
\nonumber\\
&G^0H_1H_2H_3:&\quad&
\frac{1}{v^4}\sum_{i,j,k} \epsilon_{ijk}e_ie_j^2 q_i
+\frac{1}{v^6}\sum_{i,j,k} \epsilon_{ijk}e_k^4 M_i^2 \\
&G^0G^0G^+G^-:&\quad&\frac{1}{2 v^4}\left[e_1^2M_1^2+e_2^2M_2^2+e_3^2M_3^2\right] \\
&H_iH_iG^+G^-:&\quad&
\frac{v^2-e_i^2}{2v^4}(e_1q_1+e_2q_2+e_3q_3)+\frac{e_i^2}{v^4}M_i^2
-\frac{e_i^2}{2v^6}(e_1^2M_1^2+e_2^2M_2^2+e_3^2M_3^2) \\
&H_iG^0G^+G^-:&\quad&\frac{1}{2 v^4}
\sum_{j,k}\epsilon_{ijk}e_je_k(M_k^2-M_j^2) \\
&H_iH_jG^+G^-:&\quad&-\frac{e_ie_j}{v^4}(e_1q_1+e_2q_2+e_3q_3-M_i^2-M_j^2)
-\frac{e_ie_j}{v^6}(e_1^2M_1^2+e_2^2M_2^2+e_3^2M_3^2) \\
&G^0G^0H^+H^-:&\quad& 
\frac{1}{2v^2}(e_1q_1+e_2q_2+e_3q_3) \\
&H_iH_iH^+H^-:&\quad&
\frac{v^2-e_i^2}{v^2}q
+\frac{e_i}{v^2}q_i
-\frac{e_i^2}{2v^4}(e_1q_1+e_2q_2+e_3q_3) \\
&H_iG^0H^+H^-:&\quad&
\frac{1}{v^2}\sum_{j,k}\epsilon_{ijk}e_jq_k \\
&H_iH_jH^+H^-:&\quad&
-\frac{2e_ie_j}{v^2}q
+\frac{1}{v^2}(e_jq_i+e_iq_j)
-\frac{e_ie_j}{v^4}(e_1q_1+e_2q_2+e_3q_3) \\
&G^0G^0H^+G^-:&\quad&\frac{1}{2v^4}(e_1f_1M_1^2
+e_2f_2M_2^2+e_3f_3M_3^2) \\
&H_iH_iH^+G^-:&\quad&
\frac{v^2-e_i^2}{2v^4}(f_1q_1+f_2q_2+f_3q_3)
-\frac{e_if_i}{v^4}M_{H^\pm}^2
+\frac{e_if_i}{v^4}M_i^2  \nonumber\\
&\quad&&
-\frac{e_i^2}{2v^6}(e_1f_1M_1^2+e_2f_2M_2^2+e_3f_3M_3^2) \\
&H_iG^0H^+G^-:&\quad&
-i\frac{f_i}{v^3}M_{H^\pm}^2+\frac{1}{v^4}\sum_i \epsilon_{ijk}e_jf_kM_k^2 \\
&H_iH_jH^+G^-:&\quad&
-\frac{e_ie_j}{v^4}(f_1q_1+f_2q_2+f_3q_3)
-\frac{e_if_j+e_jf_i}{v^4}M_{H^\pm}^2
+\frac{1}{v^4}(e_jf_iM_i^2+e_if_jM_j^2)  \nonumber\\
&\quad&&
-\frac{e_ie_j}{v^6}(e_1f_1M_1^2+e_2f_2M_2^2+e_3f_3M_3^2) \\
&G^+G^+G^-G^-:&\quad& \frac{1}{2 v^4}\left(e_1^2M_1^2+e_2^2M_2^2+e_3^2M_3^2\right) \\
&H^+H^+H^-H^-:&\quad& q \\
&H^+G^+H^-G^-:&\quad& 
\frac{1}{v^2}(e_1q_1+e_2q_2+e_3q_3)-\frac{2}{v^2}M_{H^\pm}^2
-\frac{1}{v^4}\left(e_1^2M_1^2+e_2^2M_2^2+e_3^2M_3^2\right)  \nonumber\\
&\quad&&
+\frac{1}{v^2}(M_1^2+M_2^2+M_3^2) \\
&H^+ H^+G^- G^-:&\quad& \frac{1}{2v^4}(f_1^2M_1^2+f_2^2M_2^2+f_3^2M_3^2) \\
&H^+G^+G^- G^-:&\quad& \frac{1}{v^4}(e_1f_1M_1^2+e_2f_2M_2^2+e_3f_3M_3^2)\\
&H^+H^+H^-G^-:&\quad&
\frac{1}{v^2}(f_1q_1+f_2q_2+f_3q_3)
\end{alignat}
Again, charge-conjugated vertices are related by complex conjugation, e.g.,
\begin{equation}
G^0G^0G^+H^- = (G^0G^0H^+G^-)^*,
\end{equation}
and are not listed separately.

%%%%%%%%%%%%%%%%%%%%%%%%%%%%%%%%%%%%%%%%%%%%%%%%%%%%%%%%%%%%
\section{2HDM Analysis in the Higgs Basis}
\label{HiggsBasis}
%%%%%%%%%%%%%%%%%%%%%%%%%%%%%%%%%%%%%%%%%%%%%%%%%%%%%%%%%%%

In the 2HDM67, the scalar fields in a general basis are parametrized by \eq{vevs}.  However, there is no physical meaning attached to this basis.  This means that the parameters $v_j$ and $\xi_j$ are unphysical.  Likewise, the cosines and sines of the mixing angles ($c_i$ and $s_i$) defined in \eq{Rmatrix} are also unphysical.  

There is some advantage to working in a basis that is more closely associated with physical parameters.   This motivates the introduction of the Higgs basis~\cite{Donoghue:1978cj,Georgi:1978ri,BLS,Davidson:2005cw}, 
\beq
\hcal_1=\begin{pmatrix}\hcal_1^+\\ \hcal_1^0\end{pmatrix}\equiv \frac{v_1e^{-i\xi_1} \Phi_1+v_2e^{-i\xi_2}\Phi_2}{v}\,,
\qquad\quad \hcal_2=\begin{pmatrix} \hcal_2^+\\ \hcal_2^0\end{pmatrix}\equiv\frac{-v_2e^{i\xi_2} \Phi_1+v_1e^{i\xi_1}\Phi_2}{v}
 \,,
 \eeq
 where $v\equiv (v_1^2+v_2^2)^{1/2}=2m_W/g=(246~{\rm GeV})^2$.
 In particular, note that the VEVs of the Higgs basis fields are
\beq \label{hbasis}
\langle \hcal_1^0\rangle=\frac{v}{\sqrt{2}}\,,\qquad\quad \langle \hcal_2^0\rangle=0\,.
\eeq
  The Higgs basis is uniquely defined up to an overall rephasing, $\hcal_2\to e^{i\chi} \hcal_2$.  
The scalar potential of the 2HDM in terms of the Higgs basis fields
is given by,
\enlargethispage{1.25\baselineskip}
\bea \mathcal{V}&=& Y_1 \hcal_1^\dagger \hcal_1+ Y_2 \hcal_2^\dagger \hcal_2 +[Y_3
\hcal_1^\dagger \hcal_2+{\rm H.c.}]
\nn\\
&&\quad 
+\half Z_1(\hcal_1^\dagger \hcal_1)^2+\half Z_2(\hcal_2^\dagger \hcal_2)^2
+Z_3(\hcal_1^\dagger \hcal_1)(\hcal_2^\dagger \hcal_2)
+Z_4( \hcal_1^\dagger \hcal_2)(\hcal_2^\dagger \hcal_1) \nn \\
&&\quad
+\left\{\half Z_5 (\hcal_1^\dagger \hcal_2)^2 +\big[Z_6 (\hcal_1^\dagger
\hcal_1) +Z_7 (\hcal_2^\dagger \hcal_2)\big] \hcal_1^\dagger \hcal_2+{\rm
H.c.}\right\}\,.\label{higgspot}
\eea

The minimization of the scalar potential yields 
\beq \label{yz}
Y_1=-\half Z_1 v^2\,,\qquad\quad Y_3=-\half Z_6 v^2\,.
\eeq
Note that the Higgs basis scalar potential parameters $Y_3$, $Z_5$, $Z_6$ and $Z_7$ acquire a phase under $\hcal_2\to e^{i\chi} \hcal_2$,
\beq
 [Y_3, Z_6, Z_7]\to e^{-i\chi}[Y_3, Z_6, Z_7] \quad{\rm and}\quad
Z_5\to  e^{-2i\chi} Z_5\,.
\eeq
In contrast, $Y_1$, $Y_2$ and $Z_{1,2,3,4}$ are invariant under $\hcal_2\to e^{i\chi} \hcal_2$.   Indeed, one can show that 
any quantity defined in the Higgs basis that is invariant under the rephasing of $\hcal_2\to e^{i\chi}\hcal_2$ is a physical quantity that is independent of the scalar basis employed to define it.\footnote{More generally,
one can show that any quantity defined in the Higgs basis
that is invariant under the rephasing of $\hcal_2\to e^{i\chi}\hcal_2$
can be rewritten explicitly in a basis-independent form. \label{fninv}}

\subsection{Identifying the scalar mass-eigenstates}

Next, we review the diagonalization of the charged Higgs and neutral Higgs squared-mass matrices. 
In the Higgs basis, the Goldstone bosons can be identified as $G^\pm=\hcal_1^\pm$ and 
$G^0=\sqrt{2}\,\Im \hcal_1^0$.    It immediately follows that the physical charged Higgs boson is $\hcal_2^\pm$, with squared mass,
\beq \label{mch}
M_{H^\pm}^2=Y_2+\half Z_3 v^2\,.
\eeq
Note that under the rephasing $\hcal_2\to e^{i\chi}\hcal_2$, the neutral and charged Goldstone fields are invariant, whereas
\beq \label{hpmshift}
\hcal^{\pm}\to e^{\pm i\chi}\hcal^{\pm}\,.
\eeq

The physical neutral Higgs bosons are linear combinations of $\sqrt{2}\,\Re \hcal_1^0-v$, $\sqrt{2}\,\Re \hcal_2^0$ and $\sqrt{2}\,\Im \hcal_2^0$.  The corresponding neutral scalar mass eigenstates are obtained by diagonalizing the $3\times 3$ real symmetric squared-mass matrix~\cite{BLS,Haber:2006ue},
\beq  \label{mtwo}
\mathcal{M}^2=v^2\left( \begin{array}{ccc}
Z_1&\,\, \Re Z_6 &\,\, -\Im Z_6\\
\Re Z_6 &\,\, \half Z_{345}+Y_2/v^2 & \,\,
- \half \Im Z_5\\ -\Im Z_6 &\,\, - \half \Im Z_5 &\,\,
 \half Z_{345}+Y_2/v^2-\Re Z_5\end{array}\right),
\eeq
where $Z_{345}\equiv Z_3+Z_4+\Re Z_5$.
The corresponding diagonalization matrix is a $3\times 3$
real orthogonal matrix that depends on three angles:
$\theta_{12}$, $\theta_{13}$ and~$\theta_{23}$,
\beq \label{mixingmatrix}
\begin{pmatrix} H_1\\ H_2 \\ H_3\end{pmatrix}
=
\begin{pmatrix} c_{12} c_{13} & \quad -s_{12}c_{23}-c_{12}s_{13}s_{23} &
\quad -c_{12}s_{13}c_{23}+s_{12}s_{23} \\
s_{12} c_{13} & \quad c_{12}c_{23}-s_{12}s_{13}s_{23} &
\quad -s_{12}s_{13}c_{23}-c_{12}s_{23} \\
s_{13} & \quad c_{13}s_{23} & c_{13}c_{23}\end{pmatrix}\begin{pmatrix}
\sqrt{2}\,\Re \hcal_1^0 -v \\ \sqrt{2}\,\Re \hcal_2^0\\ \sqrt{2}\,\Im \hcal_2^0 \end{pmatrix}\,,
\eeq
where the $H_i$ are the mass-eigenstate neutral Higgs fields,
$c_{ij}\equiv\cos\theta_{ij}$ and $s_{ij}\equiv\sin\theta_{ij}$.  Without loss of generality, the angles $\theta_{ij}$ are defined modulo $\pi$, with the convention that $c_{12}$ and $c_{13}$ are non-negative.

Under the rephasing $\hcal_2\to e^{i\chi}\hcal_2$,
\beq \label{rephasing}
\theta_{12}\,,\, \theta_{13}~{\hbox{\text{are invariant, and}}}\quad
\theta_{23}\to  \theta_{23}-\chi\,.
\eeq
As shown in Ref.~\cite{Haber:2006ue}, the invariant angles
$\theta_{12}$ and $\theta_{13}$ are basis-independent
quantities.  That is, $\theta_{12}$ and $\theta_{13}$ can be expressed explicitly in
terms of basis-independent combinations of quantities defined
in any scalar field basis [cf.~footnote~\ref{fninv}]. 

The physical neutral Higgs mass eigenstate fields are then given by,
\beq
H_k=q_{k1}\bigl(\sqrt{2}\,{\rm Re}~\hcal_1^0-v\bigr)
+\frac{1}{\sqrt{2}}\bigl(q_{k2}^*\hcal_2^0 e^{i\theta_{23}}+{\rm h.c.}\bigr)\,,
\label{hsubk}
\eeq
where the $q_{k1}$ and $q_{k2}$ are invariant
combinations of  $\theta_{12}$ and $\theta_{13}$,
which are exhibited in Table~\ref{tabinv}.
Note that the physical neutral Higgs fields, $H_j$, are manifestly invariant under the rephasing of the Higgs basis field $\hcal_2$.
%%%%%%%%%%%%%%%% BEGIN TABLE
\begin{table}[t!]
\centering
\begin{tabular}{|c||c|c|}\hline
$\phaa k\phaa $ &\phaa $q_{k1}\phaa $ & \phaa $q_{k2} \phaa $ \\
\hline
$1$ & $c_{12} c_{13}$ & $-s_{12}-ic_{12}s_{13}$ \\
$2$ & $s_{12} c_{13}$ & $c_{12}-is_{12}s_{13}$ \\
$3$ & $s_{13}$ & $ic_{13}$ \\ \hline
\end{tabular}
\caption{Invariant combinations of the neutral Higgs
boson mixing angles $\theta_{12}$ and $\theta_{13}$,
where $c_{ij}\equiv\cos\theta_{ij}$ and
$s_{ij}\equiv\sin\theta_{ij}$.
\label{tabinv}}
\end{table}
%%%%%%%%%%%%%%%% END TABLE

The following relation satisfied by the $q_{jk}$ is notable,
\beq \label{qrel}
q_{j2}^* q_{k2}=\delta_{jk}-q_{j1}q_{k1}+i\epsilon_{jk\ell}q_{\ell 1}\,.
\eeq
Setting $j=k$ then yields,
\beq \label{qsum}
q_{k1}^2+|q_{k2}|^2=1\,.
\eeq
One can also derive simple sum rules that are satisfied by the $q_{kj}$,
\beq \label{sums}
\sum_{k=1}^3 q_{k1}^2=\frac12\sum_{k=1}^3 |q_{k2}|^2=1\,,\qquad\quad
\sum_{k=1}^3 q_{k2}^2=\sum_{k=1}^3 q_{k1}q_{k2}=0\,.
\eeq
Some of these results can be understood as consequences of tree-level unitarity of the theory~\cite{Gunion:1990kf}.

One can invert \eq{hsubk} to express the Higgs basis fields $\hcal_1$ and $\hcal_2$ in terms of the mass-eigenstate scalar fields,
\beq \label{h1andh2}
\hcal_1=\begin{pmatrix} G^+ \\   
\frac{1}{\sqrt{2}}
\left[v+q_{k1}H_k+iG^0\right] \end{pmatrix}\,,
\qquad\quad
\hcal_2=  \begin{pmatrix}H^+ \\  \frac{1}{\sqrt{2}}q_{k2}e^{-i\theta_{23}} H_k \end{pmatrix}\,,
\eeq
where there is an implicit sum over the repeated index $k=1,2,3$.   
In this convention, $H^\pm\equiv\mathcal{H}_2^\pm$, which means that under the rephasing of the Higgs doublet field $\mathcal{H}_2$, the charged Higgs field acquires a phase, $H^\pm\to e^{\pm i\chi}H^\pm$ [cf.~\eq{hpmshift}], in contrast to the neutral fields $H_k$ which are invariant under the rephasing of the Higgs basis field $\hcal_2$.\footnote{This convention was employed in Ref.~\cite{Haber:2006ue}.  Alternatively, one could rephase the definition of the charged Higgs field by defining
$H^\pm\equiv e^{\pm i\theta_{23}}\hcal^\pm$, which was later adopted in Ref.~\cite{Asner:2013psa}. 
In this latter convention, all Higgs fields are invariant under the rephasing of the Higgs basis field $\hcal_2$.
Moreover, $f_k$ as defined via \eq{eflag} would now be defined in \eq{efdefs} with the factor of $e^{i\theta_{23}}$ removed, and would thus be a truly basis independent quantity.   Nevertheless, we have not adopted this alternative convention in this work.}

\subsection{Bosonic couplings of scalars and vectors in the 2HDM}

Consider the coupling of the Higgs bosons to the gauge bosons.
These arise from the Higgs boson kinetic energy terms when the
partial derivatives are replaced by the gauge covariant derivatives:
$\mathscr{L}_{\rm KE}=(D^\mu \hcal_k)^\dagger (D_\mu \hcal_k)$.
In the SU(2)$\lsub{\rm L}\times$U(1) electroweak gauge theory,
\beq \label{covder}
D_\mu \hcal_k=\left(\begin{array}{c} \displaystyle
\partial_\mu \hcal^+_k+\left[\frac{ig}{c_W}\left(\half-s_W^2\right)Z_\mu
+ieA_\mu\right] \hcal^+_k+\frac{ig}{\sqrt{2}}W_\mu^+\hcal^0_k \\[8pt]
\displaystyle \partial_\mu \hcal^0_k-\frac{ig}{2c_W}Z_\mu \hcal_k^0+
\frac{ig}{\sqrt{2}}W_\mu^- \hcal^+_k\end{array}\right)\,,
\eeq
where $s_W\equiv\sin\theta_W$ and $c_W\equiv\cos\theta_W$.  Inserting
\eq{covder} into $\mathscr{L}_{\rm KE}$ yields the Higgs boson--gauge boson
interactions in the Higgs basis.  Finally, we use \eq{h1andh2} to
obtain the interaction Lagrangian of the gauge bosons with the
physical Higgs boson mass-eigenstates and Goldstone bosons.  The resulting interaction
terms are:
\bea
\mathscr{L}_{VVH}&=&\left(gm_W W_\mu^+W^{\mu\,-}+\frac{g}{2c_W}
m_Z Z_\mu Z^\mu\right)q_{k1} H_k \nonumber \\[5pt]
&&
+em_WA^\mu(W_\mu^+G^-+W_\mu^-G^+)
-gm_Zs_W^2 Z^\mu(W_\mu^+G^-+W_\mu^-G^+)
\,,\label{VVH} \\[10pt]
\mathscr{L}_{VVHH}&=&\left[\tfrac14 g^2 W_\mu^+W^{\mu\,-}
+\frac{g^2}{8c_W^2}Z_\mu Z^\mu\right](G^0 G^0+H_k H_k) \nonumber \\
&& 
+\biggl\{
\left(\half eg A^\mu W_\mu^+ -\frac{g^2s_W^2}{2c_W}Z^\mu W_\mu^+\right)\bigl[
(q_{k1}G^-+q_{k2}\,e^{-i\theta_{23}}\,H^-)H_k+iG^-G^0\bigr] +{\rm h.c.}\biggr\}
\nonumber \\
&& +\left[\half g^2 W_\mu^+ W^{\mu\,-}+
e^2A_\mu A^\mu+\frac{g^2}{c_W^2}\left(\half
-s_W^2\right)^2Z_\mu Z^\mu \right.
\nonumber \\
&& \qquad\qquad\qquad \left.
+\frac{2ge}{c_W}\left(\half
-s_W^2\right)A_\mu Z^\mu\right](G^+G^-+H^+H^-)\,,
\label{VVHH} \\[5pt]
 \mathscr{L}_{VHH}&=&-\frac{g}{4c_W}\,\epsilon_{jk\ell}q_{\ell 1}
Z^\mu H_j\ddel_\mu H_k  
-\half g\biggl\{iW_\mu^+\left[q_{k1} G^-\ddel\lsup{\,\mu} H_k+
q_{k2}\,e^{-i\theta_{23}}\,H^-\ddel\lsup{\,\mu} H_k\right]
+{\rm h.c.}\biggr\}\nonumber \\[5pt]
&& 
+\frac{g}{2c_W} q_{k1} Z^\mu G^0\ddel_\mu H_k
+\half g\left(W_\mu^+G^-\ddel\lsup{\,\mu}G^0+W_\mu^-G^+\ddel\lsup{\,\mu}G^0
\right) \nonumber \\[5pt]
&&
+\left[ieA^\mu+\frac{ig}{c_W}\left(\half -s_W^2\right)
Z^\mu\right](G^+\ddel_\mu G^-+H^+\ddel_\mu H^-)\,, \label{VHH}
\eea
where the sum over pairs of repeated indices $j,k=1,2,3$ is implied.

The goal of this Appendix is to rewrite the cubic and quartic scalar self-coupling in terms of physical couplings and masses.   
We begin by introducing the reduced couplings $e_k$ and $f_k$ via the interaction Lagrangian,
\beq \label{eflag}
\mathscr{L}\ni\half g^2\left(W_\mu^+W^{\mu\,-}+\frac{1}{2c^2_W}
Z_\mu Z^\mu\right)e_k H_k-\frac{g}{2v}\biggl\{if^*_kW_\mu^+H^-\ddel\lsup{\,\mu} H_k
+{\rm h.c.}\biggr\}\,,
\eeq
where there is an implicit sum over the repeated index $k$.   Since the 
coupling of the scalars to vector bosons depends on the $q_{jk}$, it follows from \eqs{VVH}{VHH} that
\beq \label{efdefs}
e_k= vq_{k1}\,,\qquad f_k= vq_{k2}^*e^{i\theta_{23}}\,.
\eeq
Note that \eq{sums} yields,
\beq \label{sums2}
\sum_{k=1}^3 e_k^2=\frac12\sum_{k=1}^3 |f_k|^2=v^2\,,\qquad\quad
\sum_{k=1}^3 f_k^2=\sum_{k=1}^3 e_k f_k=0\,.
\eeq
Moreover, \eq{qsum} implies that for any choice of $k=1,2,3$,
\beq \label{ef}
e_k^2+|f_k|^2=v^2\,.
\eeq

Plugging \eq{h1andh2} into \eq{higgspot} yields
the basis-invariant form for the cubic and quartic Higgs self-couplings.  The explicit form for these scalar self-couplings can be found in Ref.~\cite{Haber:2006ue}.
In order to rewrite these couplings in terms of Higgs masses and physical couplings, we will need to introduce two additional quantities,
\beq \label{qkdef}
q_k\equiv v\bigl[q_{k1}Z_3+\Re(q_{k2}e^{-i\theta_{23}}Z_7)\bigr]\,,\qquad \quad q\equiv \half Z_2\,,
\eeq
which are defined via the scalar potential interaction terms,
\beq
\mathcal{V} \ni q_k H_k H^+ H^-+qH^+ H^- H^+ H^-\,.
\eeq
Using \eq{sums2}, it follows that 
\beq
\sum_{k=1}^3 e_k q_k=v^2 Z_3\,,\qquad\quad \sum_{k=1}^3 q^2_k=v^2\bigl(Z_3^2+|Z_7|^2\bigr)\,.
\eeq

We will also need expressions for the neutral Higgs masses.  As shown in eqs.~(C12)--(C14) of Ref.~\cite{Haber:2006ue}, one can express the squared masses of the neutral Higgs bosons in terms of $Z_1$, $Z_6$ and the neutral Higgs mixing angles.  These expressions can be compactly summarized by one equation,
\beq
M_k^2=v^2\left[Z_1+\frac{1}{q_{k1}}\Re\bigl(q_{k2}Z_6 e^{-i\theta_{23}})\right]\,,
\eeq 
where $M_k$ is the mass of $H_k$.  Using \eq{sums}, it follows that
\beq \label{weightedsum}
\sum_{k=1}^3 M_k^2 q_{k1}^2=v^2 Z_1\,.
\eeq
For completeness, we note that due to the invariance of the trace under matrix diagonalization, it follows from \eq{mtwo} that
\beq \label{mtrace}
M_1^2+M_2^2+M_3^2=2Y_2+(Z_1+Z_3+Z_4)v^2\,.
\eeq

One other mass relation that will prove useful is
\beq \label{massrel}
M_k^2-M_{H^\pm}^2=\half v^2\left[Z_4+\frac{q_{k2}}{q_{k2}^*}Z_5 e^{-2i\theta_{23}}+\frac{2q_{k1}}{q_{k2}^*}Z_6 e^{-i\theta_{23}}\right]\,.
\eeq
Since the left hand side of \eq{massrel} is manifestly real, it follows that the right hand side must be real as well.  Indeed, using eqs.~(C.8) and (C.18) of Ref.~\cite{Haber:2006ue} to eliminate $Z_5$, one can independently verify that
\beq
\Im\bigl(q_{k2}^2Z_5 e^{-2i\theta_{23}}+2q_{k1}q_{k2}Z_6 e^{-i\theta_{23}}\bigr)=0\,.
\eeq
Thus, one can take the real part of \eq{massrel} to obtain,
\beq
M_k^2-M_{H^\pm}^2=\half v^2\left[Z_4+\frac{1}{|q_{k2}|^2}\Re(q_{k2}^2 Z_5 e^{-2i\theta_{23}})+\frac{2q_{k1}}{|q_{k2}|^2}\Re(q_{k2}Z_6 e^{-i\theta_{23}})\right]\,.
\eeq

Using the above results, we can now rewrite the scalar self-couplings in terms of $e_k$, $f_k$, $q_k$, $q$ and the scalar squared-masses.  Here, we present two explicit examples.  First, we consider the cubic interaction of physical neutral Higgs scalars that arises from the scalar potential.  Using the results above, we obtain
\beq \label{Vhhh}
\mathcal{V}_{HHH}=\biggl\{\left(\delta_{k\ell}-\frac{e_k e_\ell}{v^2}\right)\left[\half q_j-\left(\frac{M_{H^\pm}^2-M_\ell^2}{v^2}\right)e_j\right]+\frac{M_\ell^2}{2v^4}e_j e_k e_\ell
\biggr\}H_j H_k H_\ell\,,
\eeq
where there are implicit sums over the repeated indices (including the index $\ell$, which is repeated three times in terms that are proportional to the squared mass $M_\ell^2$).
One can check that \eq{Vhhh} yields the results exhibited in eqs.~(\ref{hhh1}), (\ref{hhh2}) and (\ref{hhh3}).

A similar computation yields the quartic interaction of physical neutral Higgs scalars that arises from the scalar potential,
\bea
\mathcal{V}_{HHHH}&=&\biggl\{\frac{1}{4v^2}\left(\delta_{\ell m}-\frac{e_\ell e_m}{v^2}\right)\left[e_j\left(2q_k-\frac{(e_i q_i)e_k}{v^2}\right)-\frac{2e_j e_k}{v^2}\bigl(M_{H^\pm}^2-M_m^2\bigr)\right]+\frac{e_j e_k e_\ell e_m(e_i^2 M_i^2)}{8v^8}\nonumber \\
&& \qquad+\frac{q}{4}\left(\delta_{j\ell}-\frac{e_je_\ell}{v^2}\right)\left(\delta_{k m}-\frac{e_k e_m}{v^2}\right)\biggr\}H_j H_k H_\ell H_m\,,\label{Vhhhh}
\eea
where there are implicit sums over the repeated indices (including the index $m$, which is repeated three times in the terms that are proportional to the squared mass $M_m^2$).  Again, one can check that  \eq{Vhhh} yields the results exhibited in eqs.~(\ref{hhhh1}), (\ref{hhhh2}), (\ref{hhhh3}) and (\ref{hhhh4}).

\subsection{The alignment limit of the 2HDM}
\label{app:al}

Finally, we discuss the nature of the alignment limit, in which the tree-level couplings of one of the neutral Higgs bosons approach those of the SM Higgs boson.   Higgs alignment corresponds to the case in which a neutral Higgs mass eigenstate is aligned in field space with the direction of the Higgs vacuum expectation value.  That is,
$\sqrt{2}\,\Re \hcal_1^0-v$ is a mass eigenstate, which possesses the tree-level properties of the SM Higgs boson.  In light of the structure of the neutral Higgs squared mass matrix given in \eq{mtwo},
approximate alignment is realized in two cases:
\enlargethispage{1.25\baselineskip}
\begin{enumerate}[itemindent=0.5cm,label=(\roman*)]
\item $Y_2\gg v^2$, with all quartic scalar coupling parameters $Z_i$ held fixed\\ \hspace*{1.6cm}(this is the decoupling limit).
\item $|Z_6|\ll 1$, corresponding to approximate alignment with or without decoupling.
\end{enumerate}

In the decoupling regime corresponding to case (i) above, $H_2$, $H_3$ and $H^\pm$ are significantly heavier than $H_1$, with a characteristic squared mass of $\mathcal{O}(Y_2)$.  Hence, at energy scales below $(Y_2)^{1/2}$, one can integrate out all the heavy scalar states, resulting in an effective theory that can be identified as the Standard Model with a single Higgs doublet field.   Corrections to the SM Higgs properties in this effective theory scale as $v^2/Y_2$. 
Above the energy scale $(Y_2)^{1/2}$, all physical scalars are present and are approximately mass degenerate.
More precisely, squared mass differences between two of the heavy scalar states are of $\mathcal{O}(v^2)$.   

In case (ii) above, approximate alignment is realized with or without decoupling depending on whether condition (i) above is or is not satisfied.  In the latter scenario, all scalar squared masses are of $\mathcal{O}(v^2)$ or less.  Finally, \textit{exact} alignment corresponds to the infinite mass limit of the heavy scalars in case (i) above or $Z_6=0$ in case (ii) above.   Indeed, in the exact alignment limit, the tree-level couplings of $H_1^0=\sqrt{2}\Re\mathcal{H}_1^0-v$ are precisely those of the SM Higgs boson.

The conditions for alignment are easily ascertained by requiring that $\sqrt{2}\,\Re \hcal_1^0-v$ has the tree-level properties of the SM Higgs boson.  Using \eq{VVH}, one obtains
\begin{equation} \label{vrat}
\mathcal{R}_{VV}\equiv \frac{g_{H_1VV}}{g_{h_{\rm SM}VV}}=\frac{e_1}{v}=c_{12} c_{13}\,,\qquad \text{where $V=W$ or $Z$}\,,
\end{equation}
where $h_{\rm SM}$ is the SM Higgs boson.  Hence, the approximate alignment limit, $\mathcal{R}_{VV}\simeq 1$,
 corresponds to
$e_1\simeq v$.   Note that \eqs{sums2}{ef} then yield $e_2,e_3,|f_1|\ll v$ and $|f_2|\simeq |f_3|\simeq v$.

Since the $e_k$ and $f_k$ are determined by $\theta_{12}$ and $\theta_{13}$, the alignment conditions can also be  expressed as conditions on these two invariant mixing angles.  Using \eq{vrat}, it follows that approximate alignment is achieved when
$
s_{12}\,,\,s_{13}\ll 1\,.
$
One can obtain approximate expressions for $s_{12}$ and $s_{13}$ by employing the following results obtained in Ref.~\cite{Haber:2006ue}, which are a consequence of the diagonalization of the neutral scalar squared mass matrix [\eq{mtwo}],
\bea
Z_1 v^2&=&M_1^2 c_{12}^2 c_{13}^2+M_2^2 s_{12}^2 c_{13}^2 + M_3^2
s_{13}^2\,, \\
\Re(Z_6\,e^{-i\theta_{23}})\,v^2 &=& c_{13}s_{12}c_{12}(M_2^2-M_1^2)\,, \\
\Im(Z_6\,e^{-i\theta_{23}})\,v^2 &=& s_{13}c_{13}(c_{12}^2 M_1^2+s_{12}^2
M_2^2-M_3^2) \,,  \\
\Re(Z_5\,e^{-2i\theta_{23}})\,v^2 &=& M_1^2(s_{12}^2-c_{12}^2 s_{13}^2)+M_2^2(c_{12}^2-s_{12}^2 s_{13}^2)-M_3^2 c_{13}^2\,, \\
\Im(Z_5\,e^{-2i\theta_{23}})\,v^2 &=& 2s_{12}c_{12}s_{13}(M_2^2-M_1^2)\,. 
\eea
It then follows that in the approximate alignment limit, 
\bea
s_{12}&\simeq  &
\frac{\Re(Z_6 e^{-i\theta_{23}})v^2}{M_2^2-M_1^2}\ll 1\,, \label{s12al}\\
s_{13}&\simeq  &-\frac{\Im(Z_6 e^{-i\theta_{23}})v^2}{M_3^2-M_1^2}\ll 1\,.\label{s13al}
\eea
Note that the two scenarios (i) and/or (ii) that were invoked above to define the approximate alignment limit are consistent with \eqs{s12al}{s13al}.   In the decoupling limit, $M_1=125$~GeV is the mass of the SM-like Higgs boson, and $M^2_2$, $M_3^2\gg v^2$.   In the approximate alignment limit without decoupling, $|Z_6|\gg 1$ while all Higgs squared masses are of $\mathcal{O}(v^2)$.\footnote{More precisely, we require that $|Z_6|\ll \Delta M^2_{j1}/v^2$, where $\Delta M_{j1}^2\equiv M_j^2-M_1^2$ for $j=2,3$.}
Moreover, $Z_6=0$ corresponds to the exact alignment limit where $s_{12}=s_{13}=0$.

One additional small quantity characterizes the
approximate alignment limit,
\beq 
\Im(Z_5 e^{-2i\theta_{23}})\simeq \frac{2(M_2^2-M_1^2) s_{12}s_{13}}{v^2}
\simeq-\frac{\Im(Z_6^2 e^{-2i\theta_{23}})v^2}{M_3^2-M_1^2}\ll 1\,.
\eeq
Finally, the following mass relations in the approximate alignment limit are noteworthy,
\bea
M_1^2&\simeq &v^2\bigl[Z_1-s_{12}\Re(Z_6 e^{-i\theta_{23}})+s_{13}\Im(Z_6 e^{-i\theta_{23}})\bigr]\,,\\
M_2^2-M_3^2 & \simeq  &v^2\bigl[\Re(Z_5 e^{-2i\theta_{23}})+s_{12}\Re(Z_6 e^{-i\theta_{23}})+s_{13}\Im(Z_6 e^{-i\theta_{23}})\bigr]\,,
\label{massdiff1}\\
M_2^2-M_{H^\pm}^2 & \simeq  & \half v^2\bigl[Z_4+\Re(Z_5 e^{-2i\theta_{23}})+2s_{12}\Re(Z_6 e^{-i\theta_{23}})\bigr]\,.\label{massdiff2}
\eea

%%%%%%%%%%%%%%%% BEGIN TABLE
\begin{table}[t!]
\centering
\begin{tabular}{|c||c|c|c|}\hline
$\phaa k\phaa $ &\phaa $e_{k}\phaa $ & \phaa $f_{k} \phaa $ & \phaa $q_k\phaa$ \\
\hline
$1$ & $v$ & $-v(s_{12}-is_{13})e^{i\theta_{23}}$ & $v\bigl[Z_3-s_{12}\Re(e^{-i\theta_{23}}Z_7)+s_{13}\Im(e^{-i\theta_{23}}Z_7)\bigr]$ \\
$2$ & $vs_{12}$ & $ve^{i\theta_{23}}$ & $v\bigl[s_{12}Z_3+\Re(e^{-i\theta_{23}}Z_7)\bigr]$ \\
$3$ & $vs_{13}$ & $-ive^{i\theta_{23}}$  &  $v\bigl[s_{13}Z_3-\Im(e^{-i\theta_{23}}Z_7)\bigr]$\\  \hline
\end{tabular}
\caption{Values of $e_k$, $f_k$ and $q_k$ defined in \eqs{efdefs}{qkdef} in the approximate alignment limit.   Deviations from the alignment 
limit are treated to linear order in $s_{12}$ and $s_{13}$ [cf.~\eqs{s12al}{s13al}].
\label{tabal}}
\end{table}
%%%%%%%%%%%%%%%% END TABLE

In light of \eqs{s12al}{s13al}, it follows that in the exact alignment limit, the only nonzero values of the $q_{ij}$ are $q_{11}=q_{22}=1$ and $q_{32}=i$.  Working to first order in the small quantities $s_{12}$, $s_{13}$ in the approximate alignment limit generates nonzero values for $q_{21}\simeq s_{12}$, $q_{31}\simeq s_{13}$ and $q_{12}\simeq -s_{12}-is_{13}$ (all other corrections are higher order in $s_{12}$, $s_{13}$).
These results can be used to obtain the values of $e_i$, $f_i$ and $q_i$ in the approximate alignment limit, which are exhibited in Table~\ref{tabal}.
Thus, the exact alignment limit corresponds to $e_1=v$ (with $e_2=e_3=0$), and the leading deviation from the alignment limit is completely characterized by
the two small parameters $s_{12}=e_2/v$ and $s_{13}=e_3/v$, whose values are given by \eqs{s12al}{s13al}.  
 
In some cases, one must distinguish between the alignment limit with or without decoupling.   For example, 
in the exact alignment limit without decoupling, the coefficient of the $H_1 H_1 H_2$ operator obtained from \eq{hhh2} vanishes.\footnote{In obtaining this result, we have made use of the relations, 
$e_1=vc_{12}c_{13}=v[1+\mathcal{O}(s_{12}^2,s_{13}^2)]$
 and $e_2=vc_{13}s_{12}=vs_{12}[1+\mathcal{O}(s_{13}^2)]$.
It follows that $v^2-e_1^2=v^2(1-c_{12}^2 c_{13}^2)=v^2\mathcal{O}(s_{12}^2,s_{13}^2)$.}
In contrast, in the decoupling regime, $M^2_2$, $M_{H^\pm}^2\gg v^2$, and the expansion in the small parameters, $s_{12}$ and $s_{13}$, is organized differently.   In particular, \eqs{s12al}{s13al} together with \eqs{massdiff1}{massdiff2} imply that in the decoupling limit,
\beq \label{app:decouplim}
s_{12}M^2 \simeq v^2\Re(Z_6 e^{-i\theta_{23}})\,,\qquad\quad s_{13}M^2 \simeq -v^2\Im(Z_6 e^{-i\theta_{23}})\,,
\quad \text{for $M=M_2, M_3, M_{H^\pm}$}\,.
\eeq
Hence, in the exact alignment limit in the decoupling regime, the coefficient of the $H_1 H_1 H_2$ operator is finite and nonzero as $M_2\to\infty$.   Details of this analysis can be found below \eq{h1h1h2coeff}.

Using Table~\ref{tabal} together with
the cubic scalar couplings of Appendix~\ref{sect:cubic_couplings}, it is straightforward to obtain the results exhibited in Tables~\ref{cubic} and \ref{cubicDL}.

%%%%%%%%%%%%%%%%%%%%%%%%%%%%%%%%%%%%%%%%%%%%%%%%%%%%%%%%%%%%
\section{Details of the Yukawa couplings}
\label{Yukawa-more}
%%%%%%%%%%%%%%%%%%%%%%%%%%%%%%%%%%%%%%%%%%%%%%%%%%%%%%%%%%%%
The Yukawa couplings of the most general 2HDM are given in terms of the weak eigenstates\footnote{Here, $q_L^0$ and $\ell_L^0$ are the weak isospin quark doublet and lepton doublet, respectively. $u_R^0$, $d_R^0$ and $l_R^0$ are weak isospin up/down quark and lepton singlets, respectively.} in eq.~(72) of Ref.~\cite{Haber:2006ue} by
\bea
-\lcal_Y^\text{quarks}&=&\overline{q_L^0}\tilde{\Phi}_1\eta_1^{u,0}u_R^0+
\overline{q_L^0}\Phi_1\left(\eta_1^{d,0}\right)^\dagger d_R^0
+\overline{q_L^0}\tilde{\Phi}_2\eta_2^{u,0}u_R^0+
\overline{q_L^0}\Phi_2\left(\eta_2^{d,0}\right)^\dagger d_R^0+\text{h.c.}\nonumber\\
-\lcal_Y^\text{leptons}&=&
\overline{\ell_L^0}\Phi_1\left(\eta_1^{l,0}\right)^\dagger l_R^0
+\overline{\ell_L^0}\Phi_2\left(\eta_2^{l,0}\right)^\dagger l_R^0+\text{h.c.}
\label{Y_lept}
\eea
Here, $\Phi_{1,2}$ are the Higgs doublets, which we split into upper (charged) and lower (neutral) components as
\bea
\Phi_j=\left(
\begin{array}{c}\Phi_j^+\\ \Phi_j^0
\end{array}\right), \quad
\eea
and $\tilde{\Phi}_j=i\sigma_2\Phi_j^*$. Using eqs.~(73) and (74) of Ref.~\cite{Haber:2006ue} we arrive for quarks at eq.~(2.24) of Ref.~\cite{Haber:2010bw},
\bea
-\lcal_Y^\text{quarks}&=&
\overline{u}_L\left(\Phi_{\bar{a}}^{0}\right)^*\eta_a^uu_R
-\overline{d}_LK^\dagger\Phi_{\bar{a}}^-\eta_a^uu_R
+\overline{u}_LK\Phi_a^+\left(\eta_{\bar{a}}^d\right)^\dagger d_R
+\overline{d}_L\Phi_a^0\left(\eta_{\bar{a}}^d\right)^\dagger d_R+\text{h.c.},\nonumber\\
\eea
where barred and un-barred indices are to be summed over and $\eta_a^u$, $\eta_a^d$ are the Yukawa matrices in the mass eigenstate basis. We have now rotated into the mass eigenstates, and $K$ is the CKM-matrix. For leptons:
\bea
-\lcal_Y^\text{leptons}&=&
\overline{\nu_L^0}\Phi_a^+\left(\eta_{\bar{a}}^{l,0}\right)^\dagger l_R^0
+\overline{l_L^0}\Phi_a^0\left(\eta_{\bar{a}}^{l,0}\right)^\dagger l_R^0+\text{h.c.}
\eea
The parametrization of the Higgs doublets and extraction of the massless Goldstone fields and the physical mass-eigenstate fields shall be done in an identical way as in section~\ref{Sec:def-model}.
Next, in the fermionic eigenstate basis, we decompose these $\eta_i$-matrices into a part $\kappa$ proportional to the masses, and an orthogonal part $\rho$ \cite{Mahmoudi:2009zx}. Following the notation of eq.~(2.25) of Ref.~\cite{Haber:2010bw}, we have
\bea \label{Eq:eta-kappa-rho}
\eta_a^f&=&\kappa^f\hat{v}_a+\rho^f\hat{w}_a,
\eea
where $f=u$, $d$ or $l$.
Here
\bea
\hat{v}_j&=&\frac{v_j}{v}e^{i\xi_j},\\
\hat{w}_1&=&-\frac{v_2}{v}e^{-i\xi_2},\\
\hat{w}_2&=&\frac{v_1}{v}e^{-i\xi_1}.
\eea
Working out the mass terms of the Yukawa couplings, we make the following identifications
\bea
\kappa^u&=&\frac{\sqrt{2}}{v}\text{diag}(m_u,m_c,m_t),\\
\kappa^d&=&\frac{\sqrt{2}}{v}\text{diag}(m_d,m_s,m_b),\\
\kappa^l&=&\frac{\sqrt{2}}{v}\text{diag}(m_e,m_\mu,m_\tau).
\eea
This enables us to write down the Yukawa couplings of the physical mass eigenstates. They can be written in a more compact fashion if we introduce the notation
\bea
\tilde{\rho}^f=e^{-i(\xi_1+\xi_2)}\rho^f.\label{tilderho}
\eea
%%%%%%%%%%%%%%%%%%%%%%%%%%%%%%%%%%%%%%%%%%%%%%%%%%%%%%%%%%%%
\subsection{Yukawa couplings with $\rho$ diagonal}
\label{Yuk_rho_diag}
%%%%%%%%%%%%%%%%%%%%%%%%%%%%%%%%%%%%%%%%%%%%%%%%%%%%%%%%%%%%

If $\rho$ is diagonal, we avoid FCNC. The form of the Yukawa couplings when $\rho$ is diagonal is:
\bea
\bar{l}_kl_kH_j:& &-\frac{m_{l_k}}{v^2}e_j-\frac{1}{2\sqrt{2}v}
\left[
\left(\tilde{\rho}^l_{kk}\right)^*(1+\gamma_5)f_j^*
+\tilde{\rho}^l_{kk}(1-\gamma_5)f_j
\right], \label{llHj} \\
\bar{l}_k\nu_kH^-:& &-\frac{1}{2}\tilde{\rho}^l_{kk}(1-\gamma_5), \label{lnuHmin} \\
\bar{\nu}_kl_kH^+:& &-\frac{1}{2}\left(\tilde{\rho}^l_{kk}\right)^*(1+\gamma_5),\\
\bar{d}_kd_kH_j:& &-\frac{m_{d_k}}{v^2}e_j-\frac{1}{2\sqrt{2}v}
\left[
\left(\tilde{\rho}^d_{kk}\right)^*(1+\gamma_5)f_j^*
+\tilde{\rho}^d_{kk}(1-\gamma_5)f_j
\right],\\
\bar{u}_ku_kH_j:& &-\frac{m_{u_k}}{v^2}e_j-\frac{1}{2\sqrt{2}v}
\left[
\left(\tilde{\rho}^u_{kk}\right)^*(1-\gamma_5)f_j^*
+\tilde{\rho}^u_{kk}(1+\gamma_5)f_j
\right],\\
\bar{u}_md_kH^+:& &
\frac{1}{2} K_{mk}
\left[\left(\tilde{\rho}_{mm}^u\right)^*(1-\gamma_5)-\left(\tilde{\rho}_{kk}^d\right)^*(1+\gamma_5)
\right],\\
\bar{d}_ku_mH^-:& &\frac{1}{2} K_{mk}^*
\left[\tilde{\rho}_{mm}^u(1+\gamma_5)-\tilde{\rho}_{kk}^d(1-\gamma_5)
\right],\\
\bar{l}_kl_kG^0:& &-i\frac{m_{l_k}}{v}\gamma_5,\\
\bar{l}_k\nu_kG^-:& &-\frac{m_{l_k}}{\sqrt{2}v}(1-\gamma_5),\\
\bar{\nu}_kl_kG^+:& &-\frac{m_{l_k}}{\sqrt{2}v}(1+\gamma_5),\\
\bar{d}_kd_kG^0:& &-i\frac{m_{d_k}}{v}\gamma_5,\\
\bar{u}_ku_kG^0:& &i\frac{m_{u_k}}{v}\gamma_5,\\
\bar{u}_md_kG^+:& &K_{mk}\frac{m_{u_m}(1-\gamma_5)-m_{d_k}(1+\gamma_5)}{\sqrt{2}v},\\
\bar{d}_ku_mG^-:& &K_{mk}^*\frac{m_{u_m}(1+\gamma_5)-m_{d_k}(1-\gamma_5)}{\sqrt{2}v}.
\eea
where $K$ is the CKM matrix.
Note that some of the terms derived from the $\kappa$-matrices are proportional to $e_j$. Hence, in the AL, they vanish for $j=2,3$ ($H_2$ and $H_3$). The same is true for those terms containing $f_1$ ($H_1$).

%%%%%%%%%%%%%%%%%%%%%%%%%%%%%%%%%%%%%%%%%%%%%%%%%%%%%%%%%%%%
\subsection{Yukawa couplings with non-diagonal $\rho$}
\label{Yuk_gen}
%%%%%%%%%%%%%%%%%%%%%%%%%%%%%%%%%%%%%%%%%%%%%%%%%%%%%%%%%%%%

If $\rho$ is non-diagonal, we introduce new couplings with FCNC. Also, some couplings already listed above will change. The new and the changed ones are ($k\neq m$):
\bea
\bar{l}_kl_mH_j: & &
-\frac{1}{2\sqrt{2}v}
\left[
\left(\tilde{\rho}^l_{mk}\right)^*(1+\gamma_5)f_j^*
+\tilde{\rho}^l_{km}(1-\gamma_5)f_j
\right],\\
\bar{l}_m\nu_{l_k}H^-:& &-\frac{1}{2}\tilde{\rho}^l_{mk}(1-\gamma_5),\\
\bar{\nu}_{l_k}l_mH^+:& &-\frac{1}{2}\left(\tilde{\rho}^l_{mk}\right)^*(1+\gamma_5),\\
\bar{d}_kd_mH_j:& &-\frac{1}{2\sqrt{2}v}
\left[
\left(\tilde{\rho}^d_{mk}\right)^*(1+\gamma_5)f_j^*
+\tilde{\rho}^d_{km}(1-\gamma_5)f_j
\right],\\
\bar{u}_ku_mH_j:& &-\frac{1}{2\sqrt{2}v}
\left[
\left(\tilde{\rho}^u_{mk}\right)^*(1-\gamma_5)f_j^*
+\tilde{\rho}^u_{km}(1+\gamma_5)f_j
\right],\\
\bar{u}_md_kH^+:& &\frac{1}{2}\left\{
\left[(\tilde{\rho}^u)^\dag K\right]_{mk}(1-\gamma_5)
-\left[K(\tilde{\rho}^d)^\dag\right]_{mk}(1+\gamma_5)
\right\},
\\
\bar{d}_ku_mH^-:& &\frac{1}{2}\left\{
\left[K^\dag\tilde{\rho}^u\right]_{km}(1+\gamma_5)
-\left[\tilde{\rho}^d K^\dag\right]_{km}(1-\gamma_5)
\right\}.
\eea

%%%%%%%%%%%%%%%%%%%%%%%%%%%%%%%%%%%%%%%%%%%%%%%%%%%%%%%%%%%%
\subsection{Type~I Yukawa couplings}
\label{Yuk_Type_I}
%%%%%%%%%%%%%%%%%%%%%%%%%%%%%%%%%%%%%%%%%%%%%%%%%%%%%%%%%%%%

In Type~I, $\eta_1^{u,0}=\eta_1^{d,0}=\eta_1^{l,0}=0$, which immediately implies  $\eta_1^{u}=\eta_1^{d}=0$. This again implies
\begin{alignat}{2}
\rho^u&=-\frac{\hat{v}_1}{\hat{w}_1}\kappa^u, &\quad 
\tilde\rho^u&=\frac{v_1}{v_2}\frac{\sqrt{2}}{v}\text{diag}(m_u,m_c,m_t), \label{rhou_I} \\
\rho^d&=-\frac{\hat{v}_1}{\hat{w}_1}\kappa^d, &\quad 
\tilde\rho^d&=\frac{v_1}{v_2}\frac{\sqrt{2}}{v}\text{diag}(m_d,m_s,m_b), \label{rhod_I} \\
\rho^l&=-\frac{\hat{v}_1}{\hat{w}_1}\kappa^l, &\quad 
\tilde\rho^l&=\frac{v_1}{v_2}\frac{\sqrt{2}}{v}\text{diag}(m_e,m_\mu,m_\tau) \label{rhol_I}.
\end{alignat}
We find:
\bea
\bar{l}_kl_kH_j:& &-\frac{m_{l_k}}{v}\frac{R_{j2}+ic_\beta R_{j3}\gamma_5}{s_\beta},\\
\bar{l}_k\nu_kH^-:& &-\frac{m_{l_k}}{\sqrt{2}vt_\beta}(1-\gamma_5),\\
\bar{\nu}_kl_kH^+:& &-\frac{m_{l_k}}{\sqrt{2}vt_\beta}(1+\gamma_5),\\
\bar{d}_kd_kH_j:& &-\frac{m_{d_k}}{v}\frac{R_{j2}+ic_\beta R_{j3}\gamma_5}{s_\beta},\\
\bar{u}_ku_kH_j:& &-\frac{m_{u_k}}{v}\frac{R_{j2}-ic_\beta R_{j3}\gamma_5}{s_\beta},\\
\bar{u}_md_kH^+:& &
\frac{K_{mk}}{\sqrt{2}vt_\beta}
\left[m_{u_m}(1-\gamma_5)-m_{d_k}(1+\gamma_5)\right],\\
\bar{d}_ku_mH^-:& &\frac{K_{mk}^*}{\sqrt{2}vt_\beta}
\left[m_{u_m}(1+\gamma_5)-m_{d_k}(1-\gamma_5)\right],
\eea
where $t_\beta=\tan\beta=v_2/v_1$.

In the AL, Type~I Yukawa couplings preserve CP. Depending on the value of $\alpha_3$, either $H_2$ or $H_3$ will be CP-even, with the other odd.

%%%%%%%%%%%%%%%%%%%%%%%%%%%%%%%%%%%%%%%%%%%%%%%%%%%%%%%%%%%%
\subsubsection{Type~I Yukawa couplings in the alignment limit with $\alpha_3=0$}
%%%%%%%%%%%%%%%%%%%%%%%%%%%%%%%%%%%%%%%%%%%%%%%%%%%%%%%%%%%%
In this case, $H_2$ is CP-even, and the neutral-fermion Yukawa couplings can be written as
\begin{subequations}
\begin{alignat}{4}
&\bar{d}_kd_kH_1:&\quad &-\frac{m_{d_k}}{v},&\quad
&\bar{u}_ku_kH_1:&\quad &-\frac{m_{u_k}}{v},\\
&\bar{d}_kd_kH_2:&\quad &-\frac{m_{d_k}}{v}\frac{1}{t_\beta},&\quad
&\bar{u}_ku_kH_2:&\quad &-\frac{m_{u_k}}{v}\frac{1}{t_\beta},\\
&\bar{d}_kd_kH_3:&\quad &-\frac{m_{d_k}}{v}\frac{i\gamma_5}{t_\beta}, &\quad
&\bar{u}_ku_kH_3:&\quad &-\frac{m_{u_k}}{v}\frac{(-i\gamma_5)}{t_\beta},
\end{alignat}
\end{subequations}
and similarly for the leptonic couplings.
%%%%%%%%%%%%%%%%%%%%%%%%%%%%%%%%%%%%%%%%%%%%%%%%%%%%%%%%%%%%
\subsubsection{Type~I Yukawa couplings in the alignment limit with $\alpha_3=\pm\pi/2$}
%%%%%%%%%%%%%%%%%%%%%%%%%%%%%%%%%%%%%%%%%%%%%%%%%%%%%%%%%%%%
In this case, $H_2$ is CP-odd, and the neutral-fermion Yukawa couplings can be written as
\begin{subequations}
\begin{alignat}{4}
&\bar{d}_kd_kH_1:&\quad &-\frac{m_{d_k}}{v},&\quad
&\bar{u}_ku_kH_1:&\quad &-\frac{m_{u_k}}{v},\\
&\bar{d}_kd_kH_2:&\quad &-\frac{m_{d_k}}{v}\frac{(\pm i\gamma_5)}{t_\beta},&\quad
&\bar{u}_ku_kH_2:&\quad &-\frac{m_{u_k}}{v}\frac{(\mp i\gamma_5)}{t_\beta},\\
&\bar{d}_kd_kH_3:&\quad &-\frac{m_{d_k}}{v}\frac{(\mp 1)}{t_\beta}, &\quad
&\bar{u}_ku_kH_3:&\quad &-\frac{m_{u_k}}{v}\frac{(\mp 1)}{t_\beta},
\end{alignat}
\end{subequations}
and similarly for the leptonic couplings.
%%%%%%%%%%%%%%%%%%%%%%%%%%%%%%%%%%%%%%%%%%%%%%%%%%%%%%%%%%%%
\subsection{Type~II Yukawa couplings}
\label{Yuk_Type_II}
%%%%%%%%%%%%%%%%%%%%%%%%%%%%%%%%%%%%%%%%%%%%%%%%%%%%%%%%%%%%

In Type~II, $\eta_1^{u,0}=\eta_2^{d,0}=\eta_2^{l,0}=0$, which immediately implies  $\eta_1^{u}=\eta_2^{d}=0$. This again implies
\begin{alignat}{2}
\rho^u&=-\frac{\hat{v}_1}{\hat{w}_1}\kappa^u, &\quad 
\tilde\rho^u&=\frac{v_1}{v_2}\frac{\sqrt{2}}{v}\text{diag}(m_u,m_c,m_t), \label{rhou_II}\\
\rho^d&=-\frac{\hat{v}_2}{\hat{w}_2}\kappa^d, &\quad 
\tilde\rho^d&=-\frac{v_2}{v_1}\frac{\sqrt{2}}{v}\text{diag}(m_d,m_s,m_b), \label{rhod_II}\\
\rho^l&=-\frac{\hat{v}_2}{\hat{w}_2}\kappa^l, &\quad 
\tilde\rho^l&=-\frac{v_2}{v_1}\frac{\sqrt{2}}{v}\text{diag}(m_e,m_\mu,m_\tau) \label{rhol_II}.
\end{alignat}
We find:
\bea
\bar{l}_kl_kH_j:& &-\frac{m_{l_k}}{v}\frac{R_{j1}-is_\beta R_{j3}\gamma_5}{c_\beta},\\
\bar{l}_k\nu_kH^-:& &\frac{m_{l_k}}{\sqrt{2}v}t_\beta(1-\gamma_5),\\
\bar{\nu}_kl_kH^+:& &\frac{m_{l_k}}{\sqrt{2}v}t_\beta(1+\gamma_5),\\
\bar{d}_kd_kH_j:& &-\frac{m_{d_k}}{v}\frac{R_{j1}-is_\beta R_{j3}\gamma_5}{c_\beta},\\
\bar{u}_ku_kH_j:& &-\frac{m_{u_k}}{v}\frac{R_{j2}-ic_\beta R_{j3}\gamma_5}{s_\beta},\\
\bar{u}_md_kH^+:& &
\frac{K_{mk}}{\sqrt{2}vc_\beta s_\beta}
\left[c_\beta^2m_{u_m}(1-\gamma_5)+s_\beta^2m_{d_k}(1+\gamma_5)\right],\\
\bar{d}_ku_mH^-:& &\frac{K_{mk}^*}{\sqrt{2}vc_\beta s_\beta}
\left[c_\beta^2m_{u_m}(1+\gamma_5)+s_\beta^2m_{d_k}(1-\gamma_5)\right].
\eea

In the AL, Type~II Yukawa couplings preserve CP. Depending on the value of $\alpha_3$, either $H_2$ or $H_3$ will be CP-even, with the other odd.

%%%%%%%%%%%%%%%%%%%%%%%%%%%%%%%%%%%%%%%%%%%%%%%%%%%%%%%%%%%%
\subsubsection{Type~II Yukawa couplings in the alignment limit with $\alpha_3=0$}
%%%%%%%%%%%%%%%%%%%%%%%%%%%%%%%%%%%%%%%%%%%%%%%%%%%%%%%%%%%%
In this case, $H_2$ is CP-even, and the neutral-fermion Yukawa couplings can be written as
\begin{subequations}
\begin{alignat}{4}
&\bar{d}_kd_kH_1:&\quad &-\frac{m_{d_k}}{v},&\quad
&\bar{u}_ku_kH_1:&\quad &-\frac{m_{u_k}}{v},\\
&\bar{d}_kd_kH_2:&\quad &-\frac{m_{d_k}}{v}(-t_\beta),&\quad
&\bar{u}_ku_kH_2:&\quad &-\frac{m_{u_k}}{v}\frac{1}{t_\beta},\\
&\bar{d}_kd_kH_3:&\quad &-\frac{m_{d_k}}{v}(-i t_\beta \gamma_5), &\quad
&\bar{u}_ku_kH_3:&\quad &-\frac{m_{u_k}}{v}\frac{(-i\gamma_5)}{t_\beta},
\end{alignat}
\end{subequations}
and similarly for the leptonic couplings.
%%%%%%%%%%%%%%%%%%%%%%%%%%%%%%%%%%%%%%%%%%%%%%%%%%%%%%%%%%%%
\subsubsection{Type~II Yukawa couplings in the alignment limit with $\alpha_3=\pm\pi/2$}
%%%%%%%%%%%%%%%%%%%%%%%%%%%%%%%%%%%%%%%%%%%%%%%%%%%%%%%%%%%%
In this case, $H_2$ is CP-odd, and the neutral-fermion Yukawa couplings can be written as
\begin{subequations}
\begin{alignat}{4}
&\bar{d}_kd_kH_1:&\quad &-\frac{m_{d_k}}{v},&\quad
&\bar{u}_ku_kH_1:&\quad &-\frac{m_{u_k}}{v},\\
&\bar{d}_kd_kH_2:&\quad &-\frac{m_{d_k}}{v} (\mp i t_\beta\gamma_5),&\quad
&\bar{u}_ku_kH_2:&\quad &-\frac{m_{u_k}}{v}\frac{(\mp i\gamma_5)}{t_\beta},\\
&\bar{d}_kd_kH_3:&\quad &-\frac{m_{d_k}}{v}(\pm t_\beta), &\quad
&\bar{u}_ku_kH_3:&\quad &-\frac{m_{u_k}}{v}\frac{(\mp 1)}{t_\beta},
\end{alignat}
\end{subequations}
and similarly for the leptonic couplings.
%%%%%%%%%%%%%%%%%%%%%%%%%%%%%%%%%%%%%%%%%%%%%%%%%%%%%%%%%%%%
\subsection{Basis transformations for the fermionic sector}
\label{fj_trans}
%%%%%%%%%%%%%%%%%%%%%%%%%%%%%%%%%%%%%%%%%%%%%%%%%%%%%%%%%%%%
If we change basis by 
\beq
\left(
\begin{array}{c}{\Phi}^\prime_1\\ {\Phi}^\prime_2
\end{array}\right)
=
U
\left(
\begin{array}{c}\Phi_1\\ \Phi_2
\end{array}\right), 
\eeq
the Yukawa matrices $\eta_a^{f,0}$ will transform accordingly. We shall here work out the transformation rules of the Yukawa matrices under a $U(2)$ change of basis. More explicitly, we can write
\bea
\Phi_1&=&U_{11}^*{\Phi}^\prime_1+U_{21}^*{\Phi}^\prime_2,\\
\Phi_2&=&U_{12}^*{\Phi}^\prime_1+U_{22}^*{\Phi}^\prime_2.
\eea
From eq.~(\ref{Y_lept}) we find for leptons 
\bea
-\lcal_Y^\text{leptons}&=&
\overline{\ell_L^0}\left[{\Phi}^\prime_1\left(U_{11}^*\left(\eta_1^{l,0}\right)^\dagger+U_{12}^*\left(\eta_2^{l,0}\right)^\dagger\right)+{\Phi}^\prime_2\left(U_{21}^*\left(\eta_1^{l,0}\right)^\dagger+U_{22}^*\left(\eta_2^{l,0}\right)^\dagger\right)\right] l_R^0\nonumber\\
&&+\text{h.c.},
\eea
therefore 
\bea
\bigl({\eta}_1^{l,0}\bigr)^\prime&=&U_{11}\eta_1^{l,0}+U_{12}\eta_2^{l,0},\\
\bigl({\eta}_2^{l,0}\bigr)^\prime&=&U_{21}\eta_1^{l,0}+U_{22}\eta_2^{l,0}.
\eea
 The matrices $\kappa$ and $\rho$ are expressed in terms of $\eta_a$-matrices and the VEVs as follows
 \bea
 \kappa^l&=&\eta_1^{l,0}\hat{w}_2-\eta_2^{l,0}\hat{w}_1,\\
 \rho^l&=&-\eta_1^{l,0}\hat{v}_2+\eta_2^{l,0}\hat{v}_1.
 \eea
Since $\hat{v}_i$ transform in the same way as the doublets themselves therefore one finds eventually that
\bea
\bigl({\kappa}^l\bigr)^\prime&=&\kappa^l,\\
\bigl({\rho}^l\bigr)^\prime&=&\rho^l \det(U).
\eea
Since $U$ is unitary, $\det(U)$ is just a phase factor.
The same transformation rules apply to the down-quark Yukawa matrices, i.e.,
\bea
\bigl({\kappa}^d\bigr)^\prime&=&\kappa^d,\\
\bigl({\rho}^d\bigr)^\prime&=&\rho^d \det(U).
\eea
The analysis of the up-quark Yukawa matrices also yield the same transformation rules,
\bea
\bigl({\kappa}^u\bigr)^\prime&=&\kappa^u,\\
\bigl({\rho}^u\bigr)^\prime&=&\rho^u \det(U).
\eea
These transformation rules are in agreement with what is given in Ref.~\cite{Haber:2010bw}.

Since the $\kappa$-matrices are all invariant, this tells us that the fermion masses are all invariant under a change of basis (as they must be since they are observables). The $\rho$-matrices are pseudo-invariants, meaning that their absolute value is an observable.

The Yukawa couplings also contains some other quantities that are potentially sensitive to a change of basis, these are $e_i$, $f_j$, and $e^{-i(\xi_1+\xi_2)}$. We shall find out how do they transform under a U(2) basis transformation.
Let us start by illustrating the invariance of $e_i$:
\bea
{e}_j^\prime={v}_1^\prime{R}_{j1}^\prime+{v}_2^\prime{R}_{j2}^\prime=v_1R_{j1}+v_2R_{j2}=e_j,
\eea
obtained by applying the transformation rules\footnote{See Ref.~\cite{Ogreid:2017tt} for the transformation rules used here. } for $v_j$ and $R_{ij}$ under a change of basis. Next, let us consider $f_j$
\bea
{f}_j^\prime={v}_1^\prime{R}_{j2}^\prime-{v}_2^\prime{R}_{j1}^\prime-iv{R}_{j3}^\prime=e^{i\delta}\left(v_1R_{j2}-v_2R_{j1}-ivR_{j3}\right)=e^{i\delta} f_j,
\eea
where $e^{i\delta}$ is a phase factor given as
\bea
e^{i\delta}&=&\frac{\left(U_{11}v_1+U_{12}v_2e^{i(\xi_2-\xi_1)}\right)\left(U_{22}v_2+U_{21}v_1e^{-i(\xi_2-\xi_1)}\right)}{\bar{v}_1\bar{v}_2}\det[U^\dagger].\label{expidelta}
\eea
Lastly, we consider the phase factor $e^{-i(\xi_1+\xi_2)}$,
\bea
e^{-i({\xi}_1^\prime+{\xi}_2^\prime)}=e^{-i(\xi_1+\xi_2)}e^{-i({\xi}_1^\prime-\xi_1)}e^{-i({\xi}_2^\prime-\xi_2)},
\eea
where we have rewritten the expression in order to extract the factor $e^{-i(\xi_1+\xi_2)}$. In order to deal with the last two factors, we consider the transformation rules for the VEVs again to get 
\bea
e^{-i({\xi}_1^\prime-\xi_1)}&=&\frac{U_{11}^*v_1+U_{12}^*v_2e^{-i(\xi_2-\xi_1)}}{\bar{v}_1},\\
e^{-i({\xi}_2^\prime-\xi_2)}&=&\frac{U_{22}^*v_2+U_{21}^*v_1e^{i(\xi_2-\xi_1)}}{\bar{v}_2}.
\eea
This yields
\bea
e^{-i({\xi}_1^\prime+{\xi}_2^\prime)}=e^{-i(\xi_1+\xi_2)}\frac{\left(U_{11}^*v_1+U_{12}^*v_2e^{-i(\xi_2-\xi_1)}\right)\left(U_{22}^*v_2+U_{21}^*v_1e^{i(\xi_2-\xi_1)}\right)}{\bar{v}_1\bar{v}_2}.
\eea
Combining all this, we find that the combination
\bea
\bigl({\tilde{\rho}^{u,d,l}f_j}\bigr)^\prime&=&\tilde{\rho}^{u,d,l}f_j,
\eea
is invariant under a change of basis. This in turn implies that the couplings of for instance eq.~(\ref{llHj}) are invariant, hence observable. 
This implies 
\bea
\bigl({\tilde{\rho}^{u,d,l}}\bigr)^\prime&=&\tilde{\rho}^{u,d,l}e^{-i\delta},
\eea
which tells us that the couplings given in for instance eq.~(\ref{lnuHmin}) are pseudo-invariant, hence not observable (but their absolute value is observable), and they transform with the exact opposite phase as $f_j$ under a change of basis.
%%%%%%%%%%%%%%%%%%%%%%%%%%%%%%%%%%%%%%%%%%%%%%%%%%%%%%%%%%%%

%%%%%%%%%%%%%%%%%%%%%%%%%%%%%%%%%%%%%%%%%%%%%%%%%%%%%%%%%%%%
\section{Triangle functions}
\label{sect:A_B_functions}
%%%%%%%%%%%%%%%%%%%%%%%%%%%%%%%%%%%%%%%%%%%%%%%%%%%%%%%%%%%%
For the glue-glue induced Higgs production we need the scalar and pseudoscalar loop functions,
\begin{align}
A(\tau)&=2[\tau+(\tau-1)f(\tau)]\tau^{-2}, \label{Eq:A}\\
B(\tau)&=2\tau^{-1}f(\tau), \label{Eq:B}
\end{align}
with $f(\tau)$ defined by
\begin{equation}
f(\tau)=
\begin{cases}
\arcsin^2\sqrt{\tau}, & \tau\leq1\\
-\frac{1}{4}
\left[\log\frac{1+\sqrt{1-\tau^{-1}}}{1-\sqrt{1-\tau^{-1}}}-i\pi
\right]^2, &\tau>1
\end{cases}
\end{equation}
and $\tau=M_H^2/(4m_f^2)$. We recall that this function has a cusp for $\tau=1$, i.e., at the particle-antiparticle threshold. 
This cusp is present in the pseudoscalar contribution, proportional to $|B(\tau_f)|^2$, but not in the scalar one.

%%%%%%%%%%%%%%%%%%%%%%%%%%%%%%%%%%%%%%%%%%%%%%%%%%%%%%%%%%%%

\end{document}
%%%%%%%%%%%%%%%%%%%%%%%%%%%%%%%%%%%%%%%%%%%%%%%%%%%%%%%%%%%%